\documentclass[fleqn,usenatbib]{mnras}

\usepackage{newtxtext,newtxmath}
\usepackage[T1]{fontenc}

\DeclareRobustCommand{\VAN}[3]{#2}
\let\VANthebibliography\thebibliography
\def\thebibliography{\DeclareRobustCommand{\VAN}[3]{##3}\VANthebibliography}

\usepackage{graphicx}	% Including figure files
\usepackage{amsmath}	% Advanced maths commands

\newcommand{\myparagraph}[1]{\vspace{0.2em}\noindent\textbf{#1}}
\title[Probing 3D Magnetic Fields with CNN]{Probing Three-Dimensional Magnetic Fields: II - An Interpretable Convolutional Neural Network}

\author[Hu et al.]{
Yue Hu$^{1,2}$\thanks{E-mail: yue.hu@wisc.edu}
, A. Lazarian$^{2}$\thanks{E-mail: alazarian@facstaff.wisc.edu}, Yan Wu$^{3}$, Chengcheng Fu$^{4}$
\\
$^{1}$Department of Physics, University of Wisconsin-Madison, Madison, WI, 53706, USA\\
$^{2}$Department of Astronomy, University of Wisconsin-Madison, Madison, WI, 53706, USA\\
$^{3}$Computer Vision Lab, ETH Zurich, 8092, Switzerland\\
$^{4}$College of Electronics and Information Engineering, Tongji University, Shanghai, 201804, China\\
}

\date{Accepted XXX. Received YYY; in original form ZZZ}

\pubyear{2023}

\begin{document}
\label{firstpage}
\pagerange{\pageref{firstpage}--\pageref{lastpage}}
\maketitle

\begin{abstract}
Observing 3D magnetic fields, including orientation and strength, within the interstellar medium is vital but notoriously difficult. However, recent advances in our understanding of anisotropic magnetohydrodynamic (MHD) turbulence demonstrate that MHD turbulence and 3D magnetic fields leave their imprints on the intensity features of spectroscopic observations. Leveraging these theoretical frameworks, we propose a novel Convolutional Neural Network (CNN) model to extract this embedded information, enabling the probe of 3D magnetic fields. This model examines the plane-of-the-sky magnetic field orientation ($\phi$), the magnetic field's inclination angle ($\gamma$) relative to the line-of-sight, and the total magnetization level (M$_A^{-1}$) of the cloud. We train the model using synthetic emission lines of $^{13}$CO (J = 1 - 0) and C$^{18}$O (J = 1 - 0), generated from 3D MHD simulations that span conditions from sub-Alfv\'enic to super-Alfv\'enic molecular clouds. Our tests confirm that the CNN model effectively reconstructs the 3D magnetic field topology and magnetization. The median uncertainties are under $5^\circ$ for both $\phi$ and $\gamma$, and less than 0.2 for M$_A$ in sub-Alfv\'enic conditions (M$_A\approx0.5$). In super-Alfv\'enic scenarios (M$_A\approx2.0$), they are under $15^\circ$ for $\phi$ and $\gamma$, and 1.5 for M$_A$. We applied this trained CNN model to the L1478 molecular cloud. Results show a strong agreement between the CNN-predicted magnetic field orientation and that derived from Planck 353 GHz polarization. The CNN approach enabled us to construct the 3D magnetic field map for L1478, revealing a global inclination angle of $\approx76^\circ$ and a global M$_A$ of $\approx1.07$.
\end{abstract}


%%%%%%%%%%%%%%%%%%%%%%%%%%%%%%%%%%%%%%%%%%%%%%%%%%

%%%%%%%%%%%%%%%%% BODY OF PAPER %%%%%%%%%%%%%%%%%%

\section{Introduction}
In the vast interstellar medium (ISM), magnetic fields are pervasive powers that significantly influence various astrophysical phenomena. These fields serve as invisible balancers against gravitational forces within the ISM, intricately maintaining its equilibrium \citep{2018FrASS...5...39W,2020NatAs...4..704A}. They are instrumental in directing gas flows towards galactic nuclei, playing a crucial role in their sustenance and the dynamic processes unfolding therein \citep{2012ApJ...751..124K,2018MNRAS.476..235R,2020NatAs...4.1126B,2021MNRAS.506..229W,2022ApJ...941...92H}. Magnetic fields also govern the trajectories of cosmic rays, affecting the energy distribution and overall dynamics of ISM \citep{1949PhRv...75.1169F,1966ApJ...146..480J,2002PhRvL..89B1102Y,2004ApJ...614..757Y,2010A&A...510A.101F,2013ApJ...779..140X,2020ApJ...894...63X,2021MNRAS.501.4184H,2021arXiv211115066H}. Furthermore, they are deeply involved in the star formation processes within molecular clouds, influencing both the rate and nature of star births \citep{1965QJRAS...6..265M,MK04,MO07,2012ApJ...757..154L,2012ApJ...761..156F,HLS21}. Despite their pivotal roles, our understanding of these magnetic fields remains far from complete.

Our primary challenge lies in the formidable task of probing a three-dimensional (3D) magnetic field in 3D spatial space. Current approaches — such as polarized dust emission \citep{Lazarian07,BG15,2015A&A...576A.104P,2020A&A...641A..11P,2016ApJ...824..134F,2021MNRAS.tmp.3119L,2023arXiv230904173L}, polarized synchrotron emission \citep{2008A&A...482..783X,2016A&A...594A..25P,2021ApJ...920....6G}, provide 2D measurements of the plane-of-sky (POS) magnetic field direction, while Zeeman splitting \citep{Crutcher04,Crutcher12}, and Faraday rotation \citep{2007ASPC..365..242H,2009ApJ...702.1230T,2012A&A...542A..93O,2016ApJ...824..113X,2019A&A...632A..68T} provide line-of-sight (LOS) components of the magnetic field. Yielding valuable insights, these techniques probe into distinct and typically different regions of the multiphase ISM. Thus, despite their individual strengths, merging these insights into a coherent, full 3D magnetic field vector, which includes both the 3D orientation and total strength, presents a non-trivial task.

A significant advance in probing the 3D magnetic fields in molecular clouds has been made by leveraging polarized dust emission, drawing on the depolarization effect induced by different magnetic field orientations (see \citealt{2019MNRAS.485.3499C}) and by accounting for the properties of turbulent magnetic fields \citealt{2023MNRAS.519.3736H,2023MNRAS.524.4431H}). As a separate development, \cite{2019A&A...632A..68T,2022A&A...660A..97T} has succeeded in employing the synergy of Faraday rotation and dust polarization to infer a helical 3D magnetic field topology across the Orion A, Orion B, Perseus, and California clouds. Subsequently, \cite{HXL21} and \cite{HLX21a} proposed the use of anisotropic properties of magnetohydrodynamic (MHD) turbulence, inherited by young stellar objects \citep{2022ApJ...934....7H} and spectroscopic lines \citep{LP00,2016MNRAS.461.1227K,2023MNRAS.tmp.1894H}, to obtain the LOS and POS components of the magnetic field's orientation and total magnetization simultaneously. 

Importantly, the underlying theory of \cite{HLX21a}'s approach demonstrates that spectroscopic observations embody the anisotropy of MHD turbulence \citep{LP00,2016MNRAS.461.1227K,2023MNRAS.tmp.1894H}, i.e., turbulent eddies elongate along the 3D direction of the magnetic field \citep{GS95,LV99}. The spatial features present in these observations imprint the anisotropy and thus carry detailed information about the magnetic fields. This implies that, given an extensive amount of training data, machine learning algorithms have the potential to capture these features and produce accurate measurements. This strategy has been employed to map the 2D POS magnetic field orientation using velocity channel maps from spectroscopic observations \citep{2023ApJ...942...95X}. The theoretical basis remains the anisotropy of the MHD turbulence, a principle previously utilized to trace magnetic fields via velocity gradients \citep{LY18a,HYL18,2022ApJ...941...92H,2022A&A...658A..90A,2022MNRAS.510.4952L,2023MNRAS.523.1853S}. However, \cite{HLX21a} made the crucial discovery that anisotropy in velocity channel maps harbors not only information about the POS magnetic field orientation, but also the total magnetization and the magnetic field's inclination angle with respect to the LOS. This additional information paves the way for constructing the full 3D magnetic field vector from spectroscopic observations.

By leveraging the capabilities of Convolutional Neural Networks (CNN; \citealt{lecun1998gradient})—a type of deep learning model excelling in image and signal processing—we aspire to develop a novel method that can probe the 3D magnetic field. Earlier studies of the CNN explored the possibility to distinguish sub-Alfv\'enic and super-Alfv\'enic turbulence \citep{2019ApJ...882L..12P} and predict the POS magnetic field orientation \citep{2023ApJ...942...95X}. Our study, however, targets the simultaneous extraction of LOS and POS magnetic field orientations and the total magnetization. The foundation of our CNN model is the anisotropic MHD turbulence exhibited in spectroscopic observations \citep{LP00,2016MNRAS.461.1227K,2023MNRAS.tmp.1894H}, a theoretical underpinning that allows us to interpret the CNN model accurately. In other words, it enables us to discern the specific features that convey information about the magnetic field, the reasons why they are informative, and their underlying physical meanings. The effectiveness of training a CNN is highly dependent on the availability of comprehensive numerical simulations that accurately represent realistic ISM. In this research, we employ 3D MHD supersonic simulations that portray a range of ISM environments, spanning sub-Alfv\'enic scenarios (i.e., strong magnetic field), trans-Alfv\'enic, and super-Alfv\'enic conditions (i.e., weak magnetic field). We further post-process these simulations by incorporating the radiative transfer effect, which enables us to generate mock emission lines of $^{13}$CO and C$^{18}$O from diffuse molecular clouds. Through this trained CNN model, we present the 3D magnetic field map of the molecular cloud L1478. 

This paper is organized as follows. In \S~\ref{sec:theory}, we briefly review the basic concepts of MHD turbulence anisotropy in spectroscopic observations and their correlation with 3D magnetic field orientation and total magnetization. In \S~\ref{sec:data}, we give details of the 3D MHD simulations and mock observations used in this work, as well as our CNN model. We use mock observations to train the CNN model and present the results of numerical testing in \S~\ref{sec:result}. We further apply the trained CNN model to predict the 3D magnetic field in the molecular cloud L1478. In \S~\ref{sec:dis}, we discuss the uncertainty and prospects of the machine learning approach, as well as implications for various astrophysical problems. We summarize our results in \S~\ref{sec:con}.

\section{Theoretical consideration}
\label{sec:theory}
\subsection{Anisotropy of MHD turbulence: revealing magnetic field orientation and magnetization}
The earliest model of MHD turbulence was proposed to be isotropic \citep{1963AZh....40..742I,1965PhFl....8.1385K}. However, this model underwent subsequent revisions through a series of theoretical and numerical studies, revealing that MHD turbulence exhibits anisotropy under sub-Alfv\'enic conditions and isotropy at large-scale, super-Alfv\'enic conditions \citep{1981PhFl...24..825M,1983JPlPh..29..525S,1984ApJ...285..109H,1995ApJ...447..706M}.

A significant advance in this field was the introduction of the "critical balance" condition, i.e., equating the cascading time ($k_\bot v_l$)$^{-1}$ and the wave periods ($k_\parallel v_A$)$^{-1}$, proposed by \citet{GS95}, hereafter GS95. Here $k_\parallel$ and $k_\bot$ represent the components of the wavevector parallel and perpendicular to the magnetic field, respectively, while $v_l$ denotes the turbulent velocity at scale $l$, and $v_A=B/\sqrt{4\pi\rho}$ represents the Alfv\'en speed. Here $B$ is the magnetic field strength and $\rho$ is the gas mass density.

Taking into account Kolmogorov-type turbulence, i.e. $v_l\propto l^{1/3}$, the GS95 anisotropy scaling can be straightforwardly derived.
\begin{equation}
k_\parallel\propto k_\bot^{2/3},
\end{equation}
which reveals the anisotropic nature of turbulence eddies, implying that the eddies are elongated along the magnetic fields. However, it should be noted that the considerations of GS95 are based on a global reference frame, where the direction of the wavevectors is defined relative to the mean magnetic field. 
%As a consequence of the averaging effect, the dominant factor in the global frame tends to be the largest eddy, resulting in an observed anisotropy that appears scale-independent \citep{2000ApJ...539..273C}.
\begin{figure*}
	\includegraphics[width=1.0\linewidth]{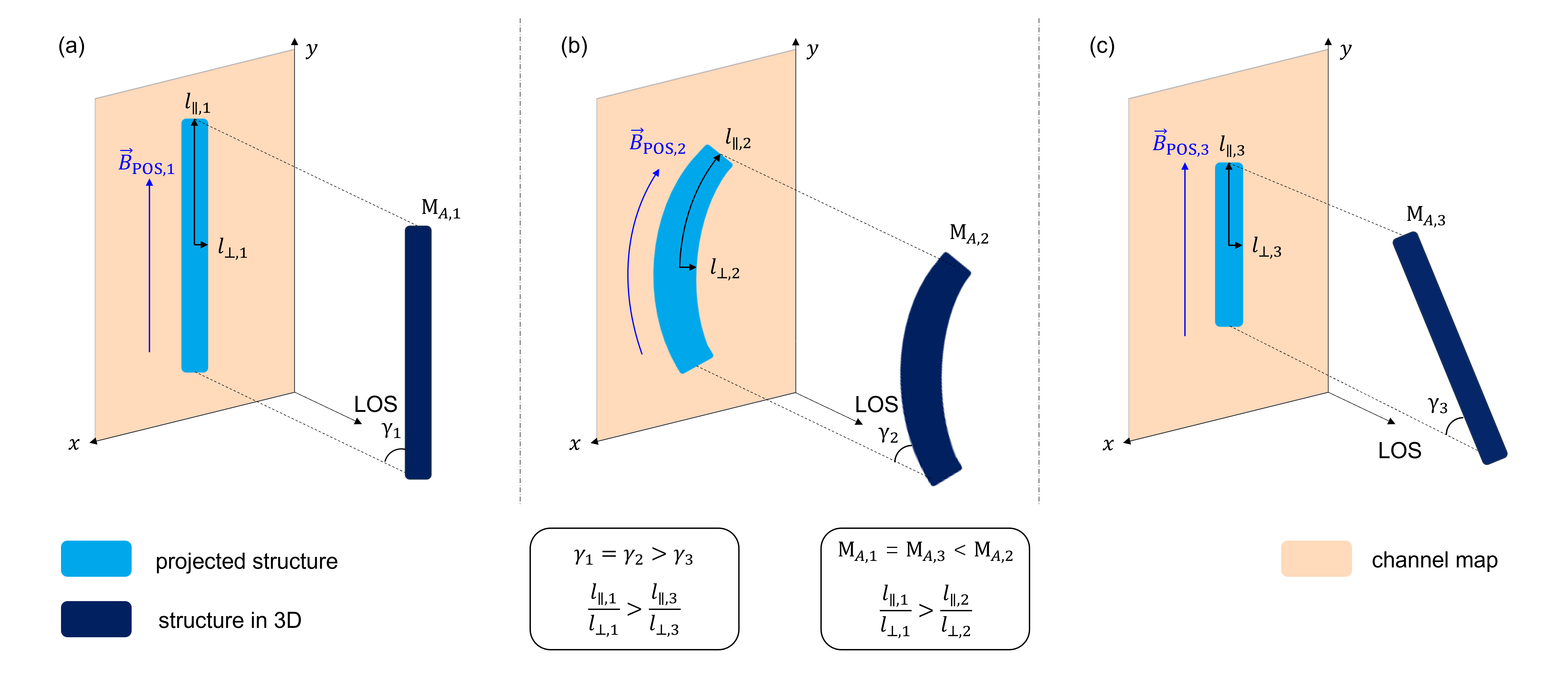}
    \caption{Illustration of how the observed intensity structures in channel map regulated by M$_A$ and $\gamma$. Within all three panels, these intensity structures elongate along the POS magnetic field direction where $l_\parallel>l_\bot$. Structures 1 and 2, depicted in panels (a) and (b), are projected onto the POS with identical inclination angles $\gamma_1=\gamma_2$, yet exhibit different magnetizations with ${\rm M}_{A,1}^{-1}>{\rm M}_{A,2}^{-1}$. Notably, the anisotropy observed, represented as $l_\parallel/l_\bot$, in the weakly magnetized Structure 2 is less pronounced than in Structure 1. Structure 2 is less straightened because the weak magnetic field has more fluctuations. The curvature of the observed magnetic structures is suggested for magnetization studies by \citet{2020ApJ...898...66Y}. Comparatively, Structures 1 and 3—showcased in panels (a) and (c)—possess equivalent magnetizations ${\rm M}_{A,1}^{-1}={\rm M}_{A,3}^{-1}$, but divergent inclination angles with $\gamma_1>\gamma_3$. The observed anisotropy decreases with smaller  $\gamma$, though it is crucial to note that the straightness of Structure 3 remains unaffected by this projection. It should be noted that here the projection effect is simplified. The intensity structures are predominantly created by the velocity caustics effect, due to MHD turbulence. The projection effect is applied to the velocity field and then subsequent intensity structures in velocity channels.}
    %anisotropy $l_\parallel/l_\bot$.}
    \label{fig:anisotropy}
\end{figure*}

Scale-dependent anisotropy was later introduced via the study of fast turbulent reconnection by \citealt{LV99} (hereafter LV99), which proposed a local reference frame. This frame is defined relative to the magnetic field passing through an eddy at scale $l$. According to LV99, the motion of eddies perpendicular to the direction of the local magnetic field adheres to the Kolmogorov law (i.e. $v_{l,\bot}\propto l_\bot^{1/3}$), since this is the direction in which the magnetic field offers minimal resistance. Applying the "critical balance" condition in the local reference frame: $v_{l,\bot}l_\bot^{-1}\approx v_Al_\parallel^{-1}$, the scale-dependent anisotropy scaling is then given by:
\begin{equation}
\label{eq.lv99}
 l_\parallel= L_{\rm inj}(\frac{l_\bot}{L_{\rm inj}})^{\frac{2}{3}}{\rm M}_A^{-4/3},~~~{\rm M}_A\le 1,
 \end{equation}
where $l_\bot$ and $l_\parallel$ represent the perpendicular and parallel scales of eddies with respect to the local magnetic field, respectively. $L_{\rm inj}$ denotes the turbulence injection scale and ${\rm M}_A=v_{\rm inj}/v_A$ is the Alfv\'en Mach number. ${\rm M}_A^{-1}$ gives magnetization level of the medium.

Eq.~\ref{eq.lv99} provides two critical insights: (1) \textbf{ Turbulent eddies stretch along the local magnetic field (i.e., $ l_\parallel\gg l_\bot$),} and (2) \textbf{the degree of anisotropy, defined as $l_\parallel/l_\bot$, depends on the magnetization ${\rm M}_A^{-1}$.} As we illustrated in Fig.~\ref{fig:anisotropy}, this indicates that eddies become increasingly anisotropic in a strongly magnetized medium. For the case where ${\rm M }_A\gg1$, turbulence is essentially isotropic due to the predominance of hydrodynamic turbulence. However, the essence of turbulence lies in the cascading of energy from larger injection scales to smaller ones, which leads to a decrease in turbulent velocity. Eventually, at the transition scale $l_a = L_{\rm inj}/M_{\rm A}^3$, the strength of the magnetic field becomes comparable to the turbulence (i.e., the Alfv\'en Mach number at $l_a$ is unity, see \citealt{2006ApJ...645L..25L}), and anisotropy starts to manifest. 

Furthermore, \textbf{(3) changes in ${\rm M}_A$ are distinctly reflected in the magnetic field topology.} Within a strongly magnetized medium, the magnetic field lines exhibit minimal variation due to the presence of weaker fluctuations, resulting in more straightened field lines. In contrast, in the context of a weaker magnetic field, which corresponds to a larger value of ${\rm M}_A$, fluctuations in the magnetic field direction intensify significantly. This leads to the field lines adopting a more curved configuration \citep{2020ApJ...898...66Y}. As turbulent eddies extend along the local magnetic field, the topological changes induced by ${\rm M}_A$ become evidently imprinted within these eddies.

\subsection{Obtaining velocity information from spectroscopic observation}
%The anisotropy of MHD turbulence leaves an imprint on the distribution of turbulent velocity fluctuations as well. Combining Eq.~\ref{eq.lv99} with the "critical balance", \citet{HXL21} and \citet{XH21} established that at a given scale $l_\parallel$, the maximum amplitude of turbulence velocity, $v_\bot$, occurs in the direction perpendicular to the local magnetic fields, whereas the minimum amplitude, $v_\parallel$, arises in the parallel direction. Specifically, the anisotropy of turbulent velocity—expressed as the ratio $v_\bot^2/v_\parallel^2$—is found to follow a power-law relation with $\rm M_A$:
%\begin{equation}
%\label{eq.vv}
%v_\bot^2/v_\parallel^2=
%\begin{cases}
%(l_\parallel/L_{\rm inj})^{-1/3}{\rm M}_A^{-4/3}, &{\rm (local, M}_A\le1)\\
%{\rm M}_A^{-4/3}, &{\rm (global,~ M}_A\le1)\\
%\end{cases}
%\end{equation}
%Here, the term "local" indicates that measurements are taken in the local reference frame, while "global" refers to measurements made in the global reference frame. 

The anisotropy outlined in Eq.~\ref{eq.lv99} pertains to turbulent velocity fluctuations, and the turbulent eddy refers to velocity fluctuation contour. This suggests that anisotropy manifests in turbulent velocity fields. Such anisotropic velocity can be obtained from the velocity channel map of spectroscopic observations, due to the velocity caustics effect \citep{LP00}. We briefly review this concept.

In position-position-velocity (PPV) space, the observed intensity distribution of a given spectral line is determined by both the density of emitters and their velocity distribution along the LOS. If coherent velocity shear — for instance, from galactic rotation — can be disregarded \footnote{The impact of galactic rotation on velocity caustics was explored by \citet{LP00}. It demonstrated that its effects are insignficant \citep{2023MNRAS.tmp.1894H}.}, the LOS velocity component, $v$, becomes the sum of the turbulent velocity, $v_{\rm tur}(x,y,z)$, and the residual component attributable to thermal motions. This residual thermal velocity, $v-v_{\rm tur}(x,y,z)$, has a Maxwellian distribution, $\phi(v,x,y,z)$. For emissivity proportional to density, it provides PPV emission density $\rho_s(x,y,z)$ as \citep{LP04}:
\begin{align}
\label{eq.max}
\rho_s(x,y,v)&=\kappa \int \rho(x,y,z) \phi(v,x,y,z) dz,\\
\phi(v,x,y,z) & \equiv \frac{1}{\sqrt{2\pi c_s^2}}\exp[-\frac{[v-v_{\rm tur}(x,y,z)]^2}{2c_s^2}],
\end{align}
where $\kappa$ is a constant that correlates the number of emitters to the observed intensities. $c_s=\sqrt{\gamma_a k_{\rm B}T/m}$ is the sound speed, with $m$ being the mass of atoms or molecules, $\gamma_a$ the adiabatic index, $k_{\rm B}$ being the Boltzmann constant, and $T$ the temperature, which can vary from point to point if the emitter is not isothermal. However, the variation of temperature has only a marginal contribution to the distribution of $\rho_s(x,y,v)$ (see \citealt{2023MNRAS.tmp.1894H}). By integrating $\rho_s(x,y,v)$ over a defined velocity range or channel width $\Delta v$, we obtain a velocity channel:
\begin{equation}
\label{eq.p}
p(x,y,v)=\int_{v-\Delta v/2}^{v+\Delta v/2}\rho_s(x,y,v^\prime)dv^\prime.
\end{equation}

By separating the 3D density into the mean density and zero-mean fluctuations, $\rho(x,y,z) = \bar\rho + \bar\rho \delta(x,y,z)$, the channel intensity can be represented as the sum of two terms, $p(x,y,v)=p_{vc}(x,y,v)+p_{dc}(x,y,v)$\citep{2023MNRAS.tmp.1894H}:
\begin{align}
\label{eq:rhov}
p_{vc}& \equiv \int_{v-\Delta v/2}^{v+\Delta v/2} \!\! dv^\prime \int \bar\rho \phi(v^\prime,x,y,z) dz, \\
\label{eq:rhod}
p_{dc}& \equiv \int_{v-\Delta v/2}^{v+\Delta v/2} \!\! dv^\prime \int \bar\rho\delta(x,y,z)\phi(v^\prime,x,y,z) dz.
\end{align}
The first term, $p_{vc}$, encompasses the mean intensity in the channel and carries fluctuations exclusively produced by velocity, called the velocity caustics effect
\citep{LP00}. The second term, $p_{dc}$, reflects the inhomogeneities in the real 3D density.

The relative importance of $p_{vc}$ and $p_{dc}$ depends on the channel width \citep{LP00,2016MNRAS.461.1227K,2023MNRAS.tmp.1894H}. The narrower the channel width, the greater the contribution from $p_{vc}$. When the channel width $\Delta v$ is less than the velocity dispersion $\sqrt{\delta (v^2)}$ of the turbulent eddies under investigation, that is, $\Delta v < \sqrt{\delta (v^2)}$, the intensity fluctuation in such a thin channel is predominantly due to velocity fluctuation. Consequently, $p(x,y,v)$ inherits the anisotropy of MHD turbulence. The intensity structures within $p(x,y,v)$ elongate along the POS magnetic fields, and their corresponding anisotropy degree, as well as the topology, is correlated with the magnetization and inclination angle. On the other hand, the dominance of $p_{vc}$ ensures that the morphology of intensity fluctuation within $p(x,y,v)$ is less sensitive to M$_s$, because the anisotropy in MHD turbulence's velocity field is not affected by M$_s$ \citep{2010ApJ...720..742K}.

It is important to note that \cite{Clark19} questioned the validity of velocity caustics in the presence of thermal broadening in multiphase HI gas and suggested that the thin velocity channel is dominated by density fluctuations from cold filaments. The nature of the striations in channel maps was tested in \cite{2023MNRAS.tmp.1894H}, by explicitly evaluating velocity and density contributions in velocity channels obtained from multi-phase HI simulations and GALFA-HI observations. This study confirmed that the velocity caustics were responsible for the observed striation.

\subsection{Anisotropy in thin velocity channels: dependence on the inclination angle of magnetic fields}
The anisotropy of the observed intensity in a PPV channel, represented by $p(x,y,v)$, is also affected by the inclination angle $\gamma$ of the magnetic field with respect to the LOS, due to the projection effect \citep{HLX21a}. For example, as illustrated in Fig.~\ref{fig:anisotropy}, we consider two magnetized structures (or eddies), $s_1$, and $s_3$, both having identical magnetization. Although these unprojected structures have the same anisotropy degree, their projections differ. Specifically, a projection with a smaller inclination angle results in a lower anisotropy degree by reducing the scale parallel to the magnetic fields. When $\gamma=0$, the parallel scale of the eddy aligns with the LOS, making the anisotropy unobservable on the POS.

However, as previously mentioned, the degree of anisotropy is also controlled by magnetization. As shown in Fig.~\ref{fig:anisotropy}, although two magnetized structures ($s_1$ and $s_2$) share identical inclination angles, the projection of the weakly magnetized $s_2$ shows less anisotropy. Importantly, the topology of $s_2$ is further changed being less straightened. This is because a weak magnetic field has more deviations and exhibits significant curvature in terms of its POS orientation \citep{2020ApJ...898...66Y}. Consequently, the observed structure, as well as the structure's topology, in $p(x,y,v)$ is governed by both $\rm M_A$ and $\gamma$ \citep{HLX21a}. %Notably, this change in anisotropy, caused by the angle of inclination, is distinguishable from the one induced by magnetization.

To summarize succinctly, the thin channel maps $p(x,y,v)$ from spectroscopic observations capture the anisotropy of MHD turbulence. This leads to the following important implications:
\begin{enumerate}
    \item The intensity structures in $p(x,y,v)$ align with the POS magnetic field.
    \item The degree of anisotropy observed in these intensity structures is influenced by two distinct factors: $\rm M_A$ and $\gamma$. These factors contribute to the anisotropy: (a) $\gamma$ introduces a projection effect that consequently decreases the anisotropy. (b) $\rm M_A$ defines the magnetization level of the medium. A larger $\rm M_A$ represents a weaker magnetic field, resulting in less pronounced anisotropy.
    \item Additionally, changes in $\rm M_A$ alter the topology of the magnetic field lines, as well as the observed intensity structure, manifesting itself as significant curvature.
\end{enumerate}
\begin{figure*}
\centering
\includegraphics[width=1.0\linewidth]{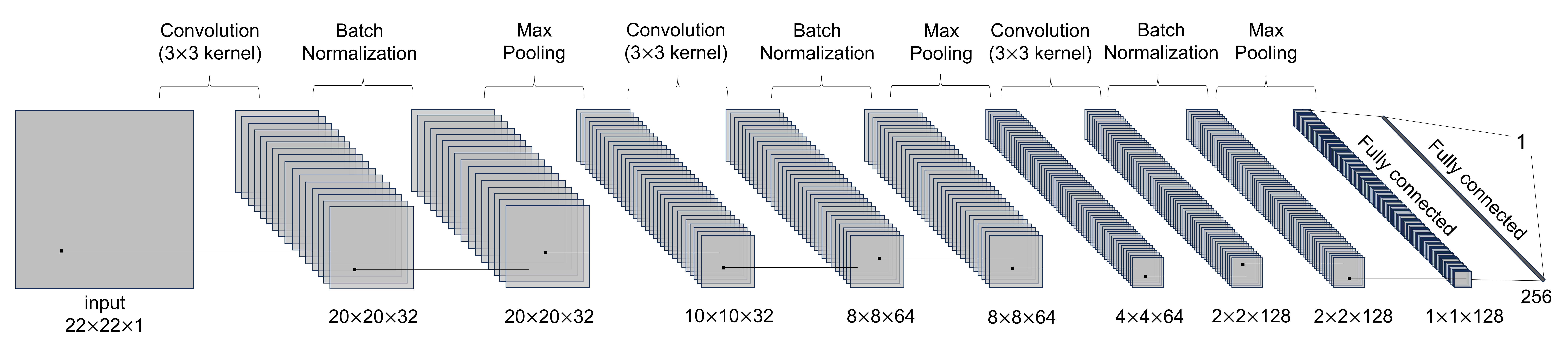}
        \caption{Architecture of the CNN-model. The input image is a $22\times22$ pixel map cropped from the thin velocity channel map. The network outputs the prediction of $\phi$, $\gamma$, or M$_A$.}
    \label{fig:cnn}
\end{figure*}

%{\bf AL: If the paper claims that it can separate the inclination angle effects from the effects of magnetization, the physical basis for this must be outlined. For instance, turbulence also induces the variation of eddy elongations and directions $\gamma$. The effects on these two quantities are different... This should be elaborated with examples. A figure could be added.}

The interconnection between magnetic field topology and $\rm M_A$ is vital to extracting accurate 3D magnetic fields. A subtle change in the degree of anisotropy responds sensitively to variations in both $\rm M_A$ and $\gamma$, leading to a degeneracy. This degeneracy necessitates the introduction of an additional feature that is sensitive to $\rm M_A$ or $\gamma$ to solve for these parameters, and the topology of the magnetic field conveniently provides this required information.

Additionally, it is crucial to acknowledge that relying solely on anisotropy does not offer a clear distinction regarding the magnetic field's orientation along the LOS, specifically whether the field is directed towards or away from our observation point. Consequently, the value of $\gamma$ is inherently restricted to a limited range between 0 and 90$^\circ$.
 
\section{Numerical method}
\label{sec:data}
\subsection{Convolutional neural network (CNN)}
To construct a deep neural network for the purpose of tracing the 3D magnetic field from a spectroscopic map, we adopt a CNN-based \citep{lecun1998gradient} architecture. CNNs have demonstrated significant success in processing multidimensional data. The typical CNN architecture, as illustrated in Fig.~\ref{fig:cnn}, consists of initial layers comprising a stack of convolutional layers followed by pooling layers. To facilitate faster convergence during the network training process using backpropagation of the loss and enhance the stability of learning, we introduce a batch normalization layer following each convolution layer. After several iterations of convolution and pooling layers, we extract a compressed image feature, which is then processed by the fully connected layers to predict the desired properties. In the following, we introduce the core modules in the CNN architecture as well as the training procedure for the CNN network.

% To build a deep neural network to estimate the magnetic field orientation, inclination angle, and total magnetization level from the image \yan{(Yue, please rephrase it, be correct and more professional for astronomer)}, we employ the convolutional neural network (CNN) design which achieved a lot of success in processing data in the format of multiple arrays, such as images, and audio spectrograms. The typical CNN architecture is shown in Fig.~\ref{fig:cnn}, of which the first few layers are composed of a stack of \textit{convolutional layers} followed by \textit{pooling layers}. To encourage the faster convergence when training the learnable parameters with loss backpropagation, \textit{Batch Normalization layer} is introduced following the \textit{convolutional layer}. After multiple layers of convolution layer + pooling layer, a compressed image feature is extracted, and \textit{fully connect layers} are followed to process the low-dimensional feature to predict desired properties.
\myparagraph{Convolutional Layer:} Serving as the fundamental component of a CNN, the convolutional layer processes input data to produce feature maps \citep{lecun1989backpropagation}. In this layer, each neuron connects to a local region of the input feature map. This connection is achieved by applying a 2D convolutional kernel $w_l$ to the input feature map. This process can be mathematically described as follows:
% \myparagraph{Convolutional Layer:} It is the CNN core build block and outputs the processed image feature map within which each neuron is connected to a local patch {\yh YH: our input is a 22*22 map, does a local patch mean a smaller sub-map?} \yan{Yan: yes} of the input feature map:
% \begin{equation}
%     a_{l,k} = \sigma(\sum_{j=1}^Kw_{l,k,j} * h_{l-1,j} + b_{l,k}),
% \end{equation}
\begin{equation}
    a_{l} = \sigma(w_{l} * h_{l-1}+ b_{l}),
\end{equation}
where $h_{l-1}$ and $a_{l}$ are the input and output feature map for the $l-{\rm th}$ convolutional layer, respectively, and $w_l$ is the learnable convolution kernel, and $*$ indicates the convolution operation. In addition, a learnable bias $b_l$ is applied to the input feature map. To be more concrete,
\begin{equation}
    a_{l}(x,y) = \sigma(\sum_{i=-k}^{k} \sum_{j=-k}^{k} w_{l}(i, j)h_{l-1}(x-i,y-j) + b_{l}(x,y)).
\end{equation}
% where $h_{l-1} \in \mathbb{R}^{d^{\rm in} \times d^{\rm in}}$ and $a_{l} \in \mathbb{R}^{d^{\rm out} \times d^{\rm out}}$ are the input and output feature map for the $l-{\rm th}$ convolutional layer, respectively. Here, $d^{\rm in}$ and $d^{\rm out}$ denote the sizes of the input and output feature maps, respectively. A learnable bias $b_l \in \mathbb{R}^{d^{\rm out} \times d^{\rm out}}$ is applied to the input feature map.
By applying the 2D convolution kernel $w_l \in \mathbb{R}^{(2k+1) \times (2k+1)}$ to the input feature map $h_{l-1} \in \mathbb{R}^{d^{\rm in} \times d^{\rm in}}$, we yield the output feature map with size ${(d^{\rm in}-k-1) \times (d^{\rm in}-k-1)}$. Here, $d^{\rm in}$ denotes the size of the input feature map and $2k+1$ is the size of the convolution kernel. The resulting locally-weighted sum, once added to the learned bias, undergoes a non-linear transformation via the ReLU activation function $\sigma(\cdot)$.

To constrain the number of parameters that need to be learned in our network, we generally use small kernel sizes. While each layer has a limited receptive field focusing on local features through the utilization of small convolutional kernels, stacking multiple layers allows for the gradual expansion of this receptive field. Consequently, the network becomes capable of capturing global features within the image as the depth increases.

% where $h_{l-1,k}$ and $a_{l,k}$ are the input and output feature map neurons for the $l-{\rm th}$ convolutional layers respectively. {\yh YH: what is the range of j?} A shared convolution 2D kernel $w_l\in \mathbb{R}^{K \times K}$ {\yh YH: what is $K\times K$?} as well as a learnable bias $b_l$ is applied on the input feature map, and the result of the locally weighted sum with the learned bias is then passed through the non-linear activation layer $\sigma(\cdot)$\yan{Yue, which non-linear layer is used here?} {\yh YH: relu}. The processed feature map captures the local conjunctions {\yh YH: what is local conjunction?} of features from the previous layer, and with a small kernel, each layer only has a limited reception field {\yh YH: reception field is?}. With the number of layers increasing, the reception field grows so that the derived feature map captures the shift-invariant global image features {\yh YH: why do we need shift-invariant global? What is the difference between global and local features?}.

\myparagraph{Batch Normalization Layer:} it is a technique frequently utilized in neural networks, playing a pivotal role in stabilizing them and hastening the convergence of the training loss during the backpropagation process \citep{ioffe2015batch}. During each training iteration, it functions on a mini-batch of data. The layer normalizes each feature within the input data by centering its values around the mean and scaling based on the feature's standard deviation within the given batch. This normalization process is instrumental in mitigating the internal covariate shift — a phenomenon where the distribution of inputs at each layer undergoes changes during training — facilitating a more stable and efficient training process.

Following the normalization, batch normalization introduces two learnable parameters per feature: a scaling parameter and a shifting parameter. These parameters allow the network to learn the optimal scale and shift for the normalized values autonomously, providing the model with the flexibility to modify the normalization if it learns that such reversal or adjustment is beneficial for its predictive performance. These dynamic adjustments, enabled by the introduced parameters, imbue the network with a degree of adaptability, allowing it to fine-tune the transformations applied to the features as needed during the training.
% After normalizing the data, BatchNorm introduces two learnable parameters for each feature: a scale parameter and a shift parameter. These parameters allow the model to learn the optimal scale and shift for the normalized values. In other words, it allows the network to adapt and potentially undo the normalization if needed.

\myparagraph{Pooling Layer:} following the detection of local features in the input feature maps by the convolution layer, a pooling layer is typically employed to merge similar local features into a singular feature \citep{sermanet2013overfeat}. One common variant of the pooling layer is the \textit{Max Pooling Layer}. This layer works by calculating the maximum value within a local patch of neurons and then outputting this maximum value as a single neuron. Importantly, the patches of input neurons for adjacent pooling units are shifted by more than one row or column, which effectively reduces the dimensionality of the feature representation. This process imparts the network with a degree of invariance to minor shifts and distortions in the input data, as it condenses the information in the feature maps while retaining the most salient features. This reduction not only helps in making the detection of features invariant to scale and orientation changes but also enhances computational efficiency by reducing the number of parameters and computations in the network.

\myparagraph{Fully Connected Layer:} After sequential operations that involve multiple convolutional layers and aggregation, the network derives a lower-dimensional compressed image feature map. Subsequently, this 2D feature map undergoes a transformation, being flattened into a 1D vector. The fully connected layer then processes this vector \citep{goodfellow2016deep}. The role of the layer is critical, as it integrates the high-level reasoning of the features extracted and flattened previously. The mechanism involves applying learned weights and biases to this flattened vector to predict the final output. Mathematically, this operation can be represented as:
\begin{equation}
\mathbf{y} = \sigma(\mathbf{W}\mathbf{h}+\mathbf{b}),
\end{equation}
In this equation, $\mathbf{h} \in \mathbb{R}^{d_{\rm in}}$ represents the flattened, compressed image feature vector, and $\mathbf{y} \in \mathbb{R}^{d_{\rm out}}$ symbolizes the predicted result. Here, $\mathbf{W} \in \mathbb{R}^{d_{\rm out} \times d_{\rm in}}$ and $\mathbf{b} \in \mathbb{R}^{d_{\rm out}}$ denote the learnable weights and biases for the fully connected layer, respectively. $d^{\rm out}$ represents the size of the output feature map. These weights and biases are integral to the layer’s functionality, providing the means for it to learn and adapt during the training phase, ultimately allowing for the accurate prediction of the desired output from the input images. %The fully connected layer essentially serves as a classifier at a high level, making sense of the data that has been transformed and filtered through the preceding layers of the network.

\myparagraph{Network Training:} The trainable parameters within the CNN are optimized by adhering to a conventional neural network training methodology, where the mean-squared error of the 3D magnetic field prediction serves as the training loss for backpropagation, as outlined in the seminal work by \citet{rumelhart1986learning}. During the training process, we implement a strategy designed to enrich the diversity of the training dataset and consequently enhance the generalization capabilities of the deep neural network. Specifically, this involves augmenting the input images by subjecting them to random cropping operations, resulting in smaller patches of size $22 \times 22$ cells. Such augmentation introduces variability and randomness into the training data, which is instrumental in refining the network's ability to generalize from the training data to unseen data, thereby bolstering its predictive accuracy and robustness. In total, we generated $\approx1.7\times10^7$ input $22 \times 22$-cell maps, with 20\% of them serving as a validation set, for each molecular species.

% {\yh Yan and Fu will write this part.}

\begin{table*}
	\centering
	\label{tab:sim}
	\begin{tabular}{ | c | c | c | c | c | c | c | c | c |}
		\hline
		Run & M$_s$ & M$_A$ & min\{M$_A^{\rm sub}$\} & max\{M$_A^{\rm sub}$\} & min\{M$_s^{\rm sub}$\} & max\{M$_s^{\rm sub}$\} \\ \hline \hline
		A0 & 5.33 & 0.20 & 0.03 & 0.28 & 2.97 & 7.84 \\ 
		A1 & 5.38 & 0.41 & 0.10 & 0.81 & 2.90 & 7.24 \\
		A2 & 5.40 & 0.61 & 0.21 & 1.00 & 3.15 & 7.33 \\  
		A3 & 5.20 & 0.79 & 0.29 & 1.37 & 3.10 & 6.55 \\  
		A4 & 5.23 & 0.95 & 0.30 & 1.99 & 3.00 & 7.18 \\
		A5 & 5.12 & 1.13 & 0.32 & 2.49 & 3.17 & 6.80 \\
		A6 & 5.38 & 1.09 & 0.41 & 3.37 & 3.13 & 6.96 \\
		A7 & 5.23 & 1.39 & 0.40 & 4.13 & 3.19 & 7.41 \\
		A8 & 5.16 & 1.46 & 0.39 & 4.94 & 3.21 & 6.76 \\
		A9 & 5.08 & 1.43 & 0.48 & 6.06 & 2.87 & 7.10 \\\hline
  		%B0 & 7.31 & 0.22 & & & & \\ 
		%B1 & 6.10 & 0.42 & & & &\\
		%B2 & 6.47 & 0.61 & & & &\\  
		%B3 & 6.14 & 0.82 & & & &\\  
		%B4 & 6.03 & 1.01 & & & &\\
		%B5 & 6.08 & 1.19 & & & &\\
		%B6 & 6.24 & 1.38 & & & &\\
		%B7 & 5.94 & 1.55 & & & &\\
		%B8 & 5.80 & 1.67 & & & &\\
		  %B9 & 5.55 & 1.71 & & & &\\
		%\hline
		%A10 & 0.47 & 0.15 & $480^3$ & $^{12}$CO, $^{13}$CO, C$^{18}O$\\
		%B2 & 7.14 & 0.66 & $480^3$ & $^{12}$CO, $^{13}$CO, C$^{18}O$\\
		%\hline
	\end{tabular}
	\caption{M$_s$ and M$_A$ are the sonic Mach number and the Alfv\'enic Mach number calculated from the global injection velocity, respectively. M$_A^{\rm sub}$ and M$_s^{\rm sub}$ are determined using the local velocity dispersion calculated along each LOS in a $22\times22$ cell sub-field. The expressions "min\{...\}" and "max\{...\}" denote the minimum and maximum value averaged over each $22\times22$ cell sub-field within the corresponding simulation.
 }
\end{table*}

\subsection{MHD simulations}
The numerical simulations used in this study were executed using the ZEUS-MP/3D code \citep{2006ApJS..165..188H}. We performed an isothermal simulation of a 10 pc cloud by solving the ideal MHD equations in an Eulerian frame under periodic boundary conditions:
\begin{equation}
\label{eq.mhd}
\begin{aligned}
    &\partial\rho/\partial t +\nabla\cdot(\rho\pmb{v})=0,\\
    &\partial(\rho\pmb{v})/\partial 
    t+\nabla\cdot\left[\rho\pmb{v}\pmb{v}^T+(c_s^2\rho_+\frac{B^2}{8\pi})\pmb{I}-\frac{\pmb{B}\pmb{B}^T}{4\pi}\right]=\pmb{f},\\
&\partial\pmb{B}/\partial t-\nabla\times(\pmb{v}\times\pmb{B})=0,\\
    &\nabla \cdot\pmb{B}=0,\\
    \end{aligned}
\end{equation}
where $\pmb{f}$ represents the stochastic forcing term used to drive turbulence. $\rho$, $\pmb{v}$, and $\pmb{B}$ are mass density, velocity, and magnetic field, respectively. Given the isothermal equation of state, the sound speed $c_s$ was held constant at approximately 187 m/s, corresponding to a gas temperature of 10 K. Purely turbulent scenarios were also considered, excluding the impact of self-gravity. Kinetic energy was solenoidally (i.e., the forcing term is divergence-free) injected at the wavenumber $k=2\pi/l\approx2$ (in the unit of $2\pi/L_{\rm box}$, where $L_{\rm box}$ is the length of simulation box) in Fourier space, where $l$ is the length scale in real space, producing a Kolmogorov-like power spectrum. Turbulence was continuously stimulated until it reached a state of statistical saturation. The simulation was solved on a regular grid of 792$^3$ cells and the turbulence was numerically dissipated at scales of approximately 10 - 20 cells.

The simulations were initialized with a uniform density field and a magnetic field, with the initial mean magnetic field oriented along the y-axis. Furthermore, we rotated the simulation cubes so that the mean angle of inclination with respect to the LOS (or z-axis) reached $90^\circ$, $60^\circ$, and $30^\circ$. The sonic Mach number, ${\rm M}_s=v_{\rm inj}/c_{s}$, and the Alfv\'enic Mach number, ${\rm M}_A=v_{\rm inj}/v_A$, characterize MHD turbulence simulations. To model different ISM conditions, we used a typical mean number density of ~300 cm$^{-3}$ and varied the initial uniform magnetic field and the injected kinetic energy to obtain a range of ${\rm M}_A$ and ${\rm M}_s$ values. In this paper, we refer to the simulations in Tab.~1 by their model name or key parameters.
\begin{figure*}
\includegraphics[width=1.0\linewidth]{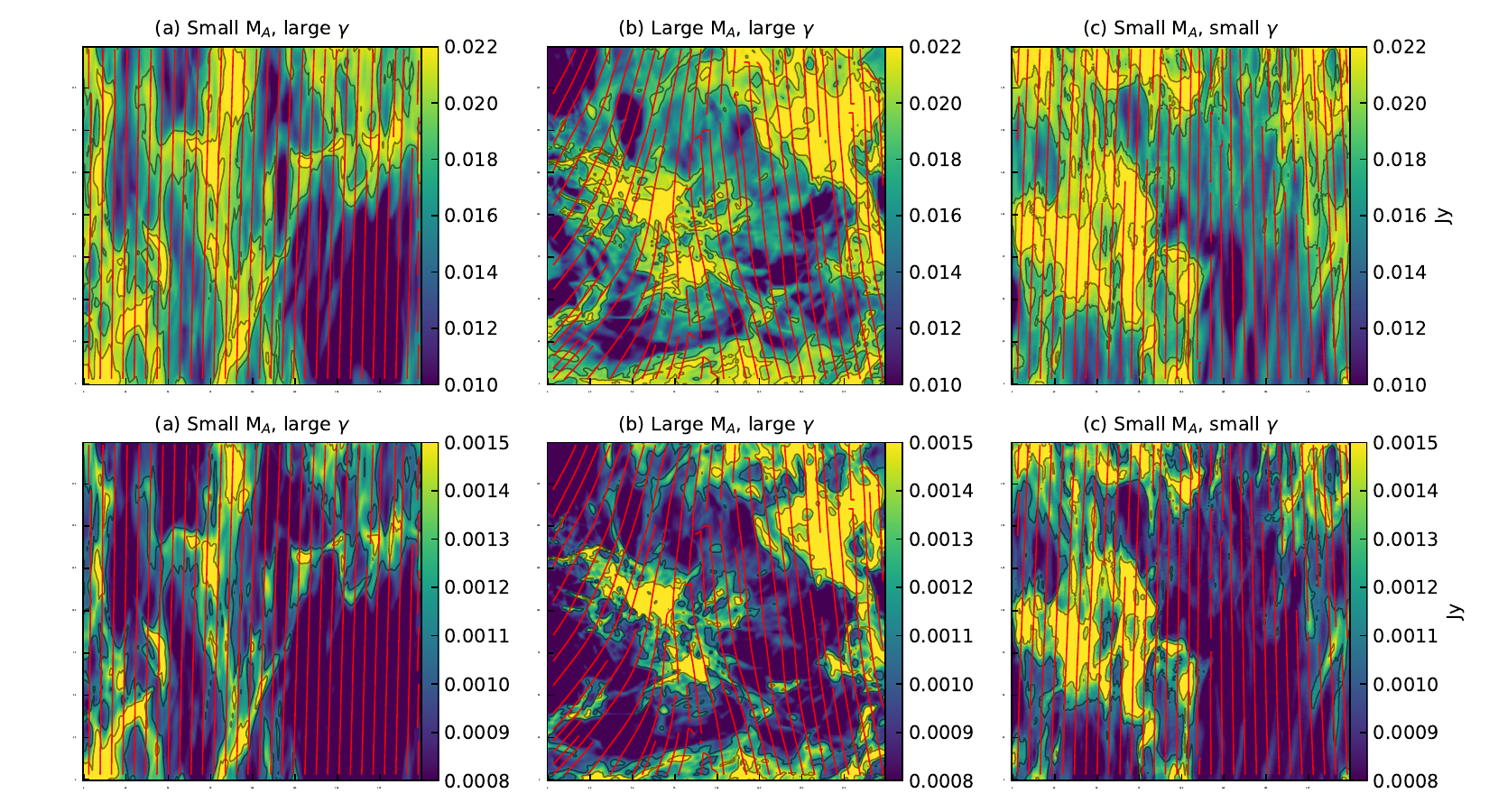}
        \caption{An numerical illustration of the anisotropy in $^{13}$CO (top) and C$^{18}$O channel map. The red streamlines represent the POS magnetic field orientation. Panel (a): M$_A=0.20$, $\gamma=90^\circ$. 
        Panel (b): M$_A=1.43$, $\gamma=90^\circ$. Panel (c): M$_A=0.20$, $\gamma=60^\circ$.}
    \label{fig:map}
\end{figure*}

\begin{figure*}
\centering
\includegraphics[width=0.85\linewidth]{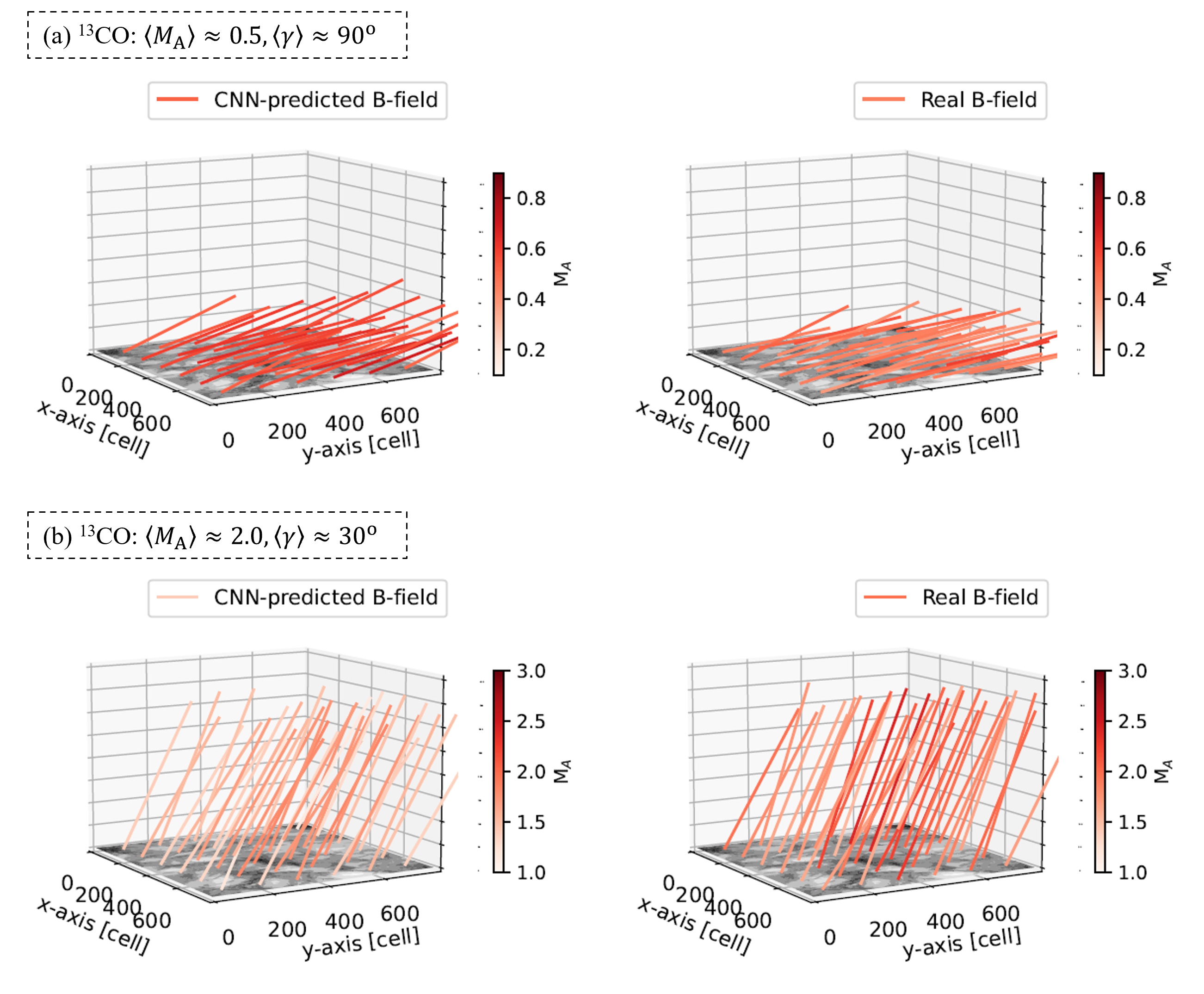}
        \caption{An comparison of the CNN-predicted 3D magnetic fields using $^{13}$CO in sub-Alfv\'en (top, $\langle {\rm M}_A\rangle\approx0.5$ and $\langle \gamma\rangle\approx90^\circ$) and super-Alfv\'en (bottom,  $\langle {\rm M}_A\rangle\approx2.0$ and $\langle \gamma\rangle\approx30^\circ$) conditions. Each magnetic field segment is constructed by the POS magnetic field's position angle (i.e., $\phi$) and the inclination angle $\gamma$. Note that the magnetic field obtained is the projection along the LOS and averaged over 132$\times$132 pixels for visualization purposes. The third axis of the LOS is for 3D visualization purposes and does not provide distance information here. The total intensity map $I$ is placed on the POS, i.e., the $x-y$ plane. }
    \label{fig:3D_13co}
\end{figure*}

\begin{figure*}
\includegraphics[width=0.85\linewidth]{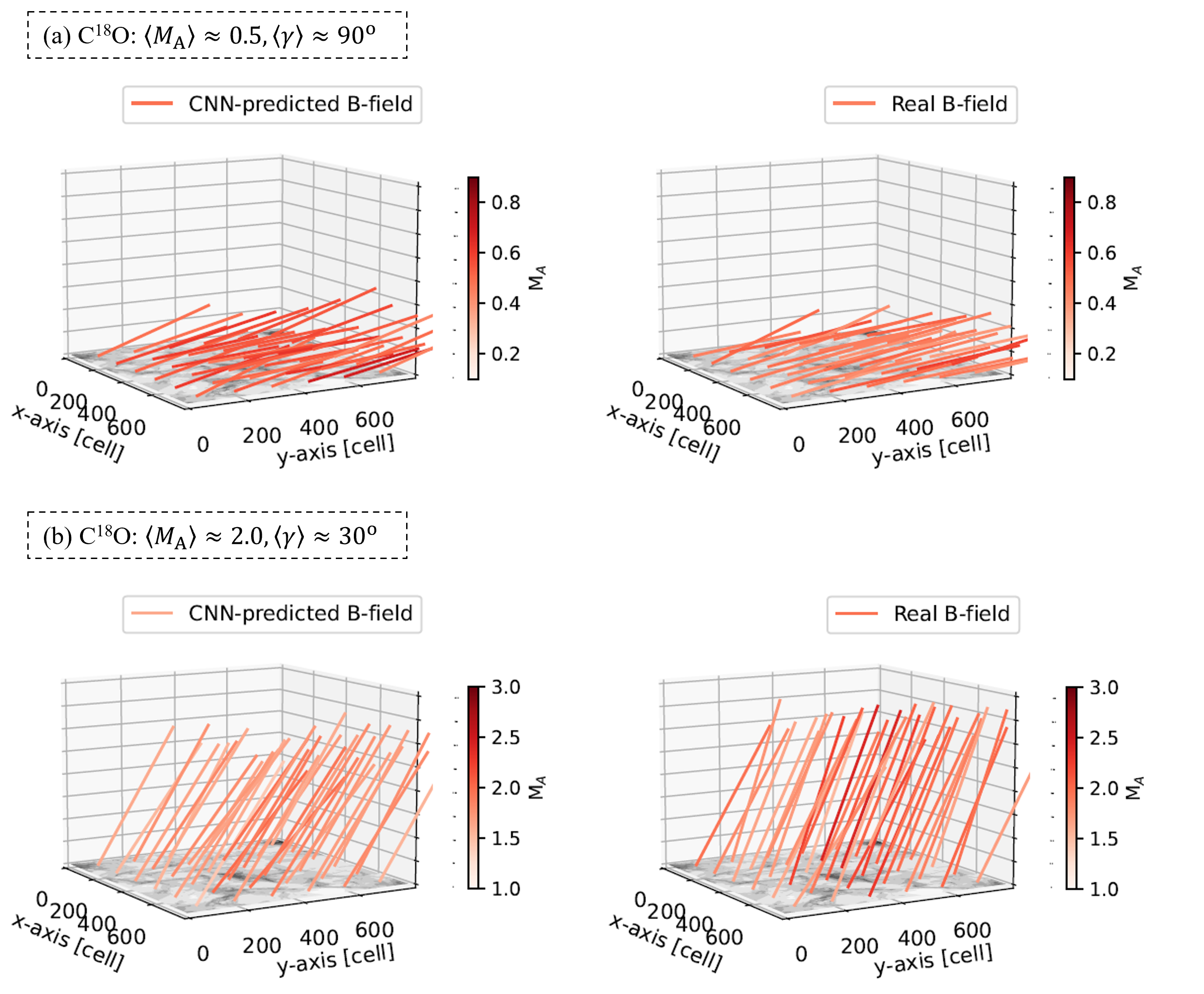}
        \caption{Same as Fig.~\ref{fig:3D_13co}, but for C$^{18}$O.}
    \label{fig:3D_c18o}
\end{figure*}

\begin{figure*}
\includegraphics[width=1.0\linewidth]{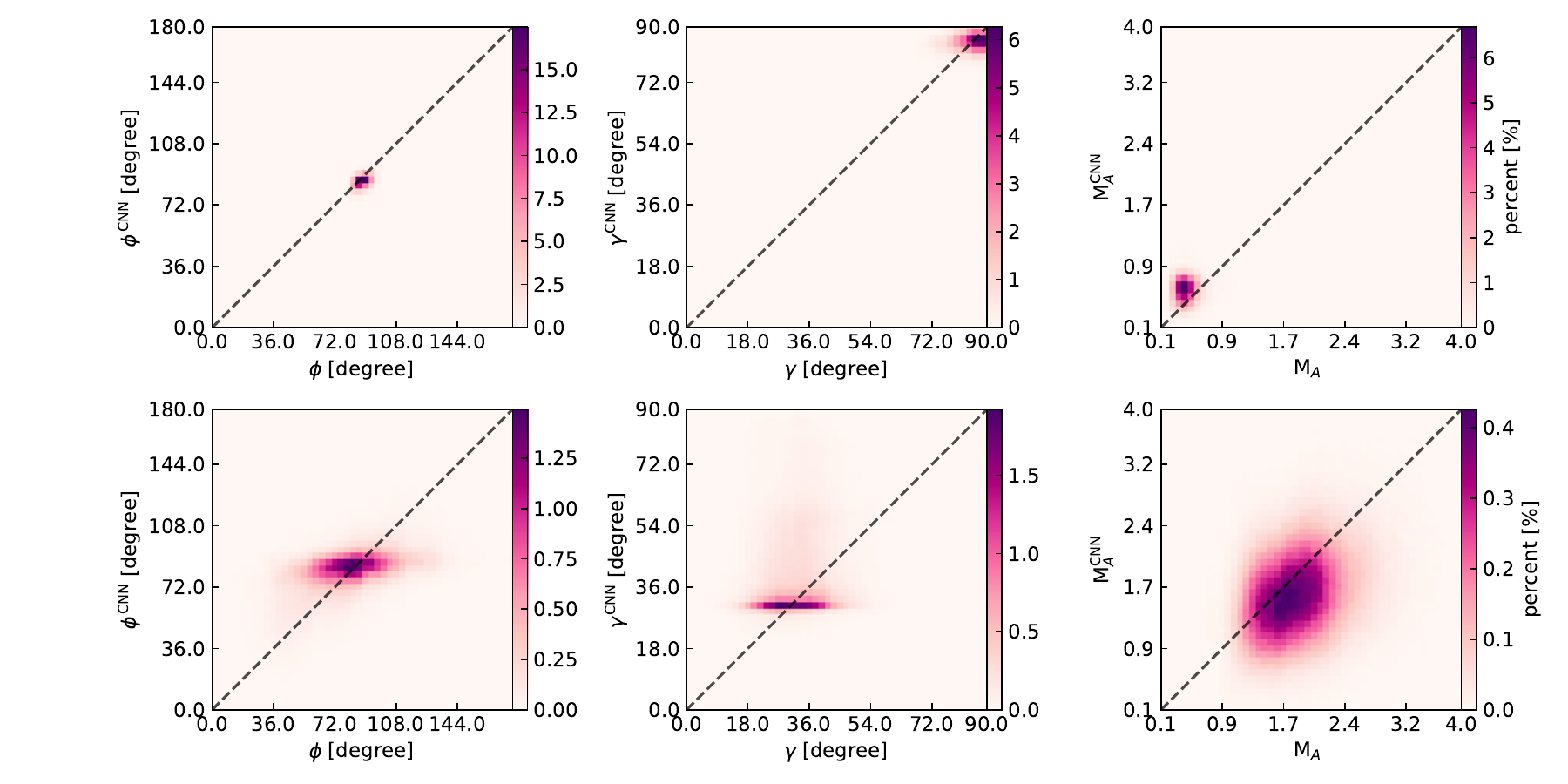}
        \caption{2D histogram of the $^{13}$CO CNN-predictions, i.e., $\phi^{\rm CNN}$ (left), $\gamma^{\rm CNN}$ (middle), and M$_A^{\rm CNN}$ (right) and the corresponding actual values in simulation (Top: sub-Alfv\'en, $\langle {\rm M}_A\rangle\approx0.5$ and $\langle \gamma\rangle\approx90^\circ$. Bottom: super-Alfv\'en, $\langle {\rm M}_A\rangle\approx2.0$ and $\langle \gamma\rangle\approx30^\circ$). The dashed reference line represents the ideal scenario, where the predicted values and actual values match perfectly.}
    \label{fig:2Dhist_13}
\end{figure*}

\subsection{Emission lines of $^{13}$CO and C$^{18}$O}
We generate synthetic emission lines for two CO isotopologues:  $^{13}$CO (1-0) and C$^{18}$O (1-0), following the procedures used in \cite{2021MNRAS.502.1768H}. This was achieved using the SPARX radiative transfer code \citep{2019ApJ...873...16H}. SPARX solves the radiative transfer equation (RTE) for finite cells, which means that it considers the emission from a homogeneous finite element. The equation of statistical equilibrium for molecular levels takes into account molecular self-emission, stimulated emission, and collisions with gas particles. Information on the distribution of molecular gas density with mean density $\sim300~{\rm cm^{-3}}$ and LOS velocity was extracted from the MHD simulations mentioned above.

The fractional abundances of the CO isotopologues $^{13}$CO(1-0) and C$^{18}$O(1-0) were set at $2\times10^{-6}$ and $1.7\times10^{-7}$, respectively.
%following \citet{2019ApJ...873...16H,2021MNRAS.502.1768H}. 
We derive the $^{12}$CO-to-H$_2$ ratio of $1\times10^{-4}$ from the cosmic value of C/H = $3\times10^{-4}$ and the assumption that 15\% of C is in molecular form. The abundance of $^{13}$ CO is determined using a $^{13}$CO/$^{12}$CO ratio of 1/69, as indicated by \citet{1999RPPh...62..143W}, giving a $^{13}$ CO / H$_2$ ratio of approximately $2\times10^{-6}$. Using a $^{12}$CO/C$^{18}$O ratio of 500, as given by \citet{2013tra..book.....W}, we obtained a C$^{18}$O-to-H$_2$ ratio of $1.7\times10^{-7}$. When generating these synthetic emission lines, we specifically focused on the lowest-transition J = 1-0 of the CO isotopologues, with the Local Thermodynamic Equilibrium (LTE) satisfied.

\subsection{Training images}
Our training input is a thin velocity channel map, $p(x,y,v_0)$, derived from either the $^{13}$CO (1-0) or C$^{18}$O (1-0) line, calculated from:
\begin{equation}
\label{eq.p}
\begin{aligned}
p(x,y,v_0)=\int_{v_0-\Delta v/2}^{v_0+\Delta v/2}T_{\rm e}(x,y,v)dv,\
\end{aligned}
\end{equation}
where $v_0$ is the velocity associated with the line's central peak, $T_{\rm e}$ is the emission line's intensity, and $\Delta v=\sqrt{\delta(v^2)}$. Here $\sqrt{\delta(v^2)}$ is the velocity dispersion derived from the moment-1 map (velocity centroid map). The $^{12}$CO line, a common diffuse cloud tracer, is not used in this work due to numerical limitations related to the saturation of the intensity of $^{12}$CO in the channel centering at $v_0$, which obliterates the spatial features of that channel \citep{2019ApJ...873...16H}. However, the CNN method could be extended to include wing channels centering at $|v|<v_0$ to bypass this numerical saturation, a possibility we might explore in future work.\footnote{The use of wing channels has its own advantages through increasing the ratio of velocity to density fluctuations \citep{2021ApJ...910..161Y,2023MNRAS.tmp.1894H}.}

We generate $p(x,y,v_0)$ for the full cloud, a region of $792\times792$ cells, then randomly segment $p(x,y,v_0)$ into $22\times22$-cell subfields for input into the CNN model. The choice of $22\times22$-cell avoids that the features fall into the numerical dissipation range, in which the anisotropy of MHD turbulence is distorted by numerical diffusivity. In observation, the inertial range of MHD turbulence is much longer and the velocity channel map is not affected by the dissipation. The size of the sub-field, thus, could be smaller to achieve higher resolution. For each subfield, we also generate corresponding projected maps of $\phi^{\rm sub}$, $\gamma^{\rm sub}$, M$_A^{\rm sub}$, and M$_s^{\rm sub}$ as per the following:
\begin{equation}
\begin{aligned}
\phi^{\rm sub}(x,y)&=\arctan(\frac{\int B_y(x,y,z)dz}{\int B_x(x,y,z)dz}),\\
\gamma^{\rm sub}(x,y)&=\arccos(\frac{\int B_z(x,y,z)dz}{\int B(x,y,z)dz}),\\
{\rm M}_A^{\rm sub}&=\frac{v_{\rm inj}^{\rm los}\sqrt{4\pi\langle\rho\rangle_{\rm los}}}{\langle B\rangle_{\rm los}}, \\
{\rm M}_s^{\rm sub}&=\frac{v_{\rm inj}^{\rm los}}{c_s},
\end{aligned}
\end{equation}
where $B=\sqrt{B_x^2+B_y^2+B_z^2}$ is the total magnetic field strength, and $B_x$, $B_y$, and $B_z$ are its $x$, $y$, and $z$ components. $\langle\rho\rangle_{\rm los}$ and $\langle B\rangle_{\rm los}$ are the gas mass density and magnetic field strength averaged along the LOS. M$_A^{\rm sub}$ and M$_s^{\rm sub}$ are defined using the local velocity dispersion for each LOS (i.e., $v_{\rm inj}^{\rm los}$), rather than the global turbulent injection velocity $v_{\rm inj}$ used to characterize the full simulation.
%making M$_A^{\rm sub}$ smaller than M$_A=v_{\rm inj}/v_A$, as $v$ is smaller than $v_{\rm inj}$ (the global turbulent injection velocity). 
The ranges of M$_A^{\rm sub}$ and M$_s^{\rm sub}$ averaged over the subfield in each simulation with different $\gamma$ are listed in Tab.~1, while $\gamma^{\rm sub}$ spans from 0 to 90$^\circ$. These values of M$_A^{\rm sub}$, M$_s^{\rm sub}$, and $\gamma^{\rm sub}$ cover typical physical conditions of diffuse molecular clouds \citep{2023MNRAS.524.4431H}.

\section{Results}
\label{sec:result}
\subsection{Numerical training and tests}
Fig.~\ref{fig:map} provides a visualization detailing the influence of M$_A$ and $\gamma$ on the anisotropy of intensity structures within thin velocity channels. In scenarios where both M$_A$ and $\gamma$ values are small, the intensity structures distinctly manifest as slender strips, extending in alignment with the POS magnetic fields. These structures are produced predominantly by the turbulent velocity \citep{LP00}, as demonstrated in \cite{2023MNRAS.tmp.1894H}. As M$_A$ increases, representing a weakening in the magnetic field, the MHD turbulence begins to more closely resemble isotropic hydrodynamical turbulence. This shift brings about a marked change in the topology of intensity structures, making them less anisotropic. Alternatively, when dealing with smaller values of $\gamma$, which imply that magnetic fields are oriented more proximally to the LOS, the inherent anisotropy is subdued due to the projection effect. Comparing $^{13}$CO and C$^{18}$O, C$^{18}$O is more sensitive to denser gas, so its associated intensity structures exhibit distinct characteristics. Despite these differences, the underlying physical principle of anisotropic MHD turbulence remains the same, suggesting M$_A$ and $\gamma$ continue to shape the observed structural formations. 

Fig.~\ref{fig:3D_13co} provides a comparative visualization between the actual 3D magnetic fields and those predicted through the utilization of the trained CNN model with $^{13}$CO. This comparison is framed within two distinct conditions: sub-Alfv\'enic (simulation with $\langle {\rm M}_A\rangle\approx0.5$ and $\langle \gamma\rangle\approx90^\circ$) and super-Alfv\'enic (simulation with $\langle {\rm M}_A\rangle\approx2.0$ and $\langle \gamma\rangle\approx30^\circ$). Within these settings, the mean projected total Alfv\'en Mach number on the POS is given as $\langle {\rm M}_A\rangle\approx0.5$ for sub-Alfv\'enic conditions and $\langle {\rm M}_A\rangle\approx2.0$ for super-Alfv\'enic ones.

The visual segment displayed in Fig.~\ref{fig:3D_13co} is constructed from the POS magnetic field’s position angle, $\phi$, and the inclination angle, $\gamma$, with a superimposed color representation signifying the projected M$_A$. Upon comparison with the intrinsic magnetic field embedded within the simulation, a noteworthy observation is the alignment between the orientations of the CNN-predicted 3D magnetic field and the actual field, evident under both sub-Alfv\'enic and super-Alfv\'enic conditions. In the sub-Alfv\'enic case, the CNN-predicted ${\rm M}_A$ is slightly larger (by $\approx 0.1-0.2$) than the actual values. Conversely, in the super-Alfv\'enic scenario, the predicted value is somewhat smaller, with a deviation ranging from $\approx 0.5-1.0$. 
Another example with
$\langle {\rm M}_A\rangle\approx0.15$ and $\langle \gamma\rangle\approx60^\circ$, is presented in Appendix~\ref{appendix.A}. Although this simulation shows an anisotropy degree similar to the case with $\langle {\rm M}_A\rangle\approx0.5$ and $\langle \gamma\rangle\approx90^\circ$, the CNN model effectively resolves the degeneracy in the correlation of the anisotropy degree with $\gamma$ and M$_A$ (see \S~\ref{sec:theory}), successfully recovering the 3D magnetic field (see Fig.~\ref{fig:3D_13cob}). It should be noted that the predicted M$_A$ is still overestimated by approximately 0.1 - 0.2.

Fig.~\ref{fig:3D_c18o} offers a similar visual comparison but focuses on the C$^{18}$O line. This line is generally recognized as denser tracers compared to $^{13}$CO. Despite these differences in tracer density, the CNN predictions for C$^{18}$O lines maintain a general alignment with the actual 3D magnetic fields observed within the simulations. Moreover, there is less significant overestimation and underestimation in the CNN-predicted ${\rm M}_A$.

\begin{figure*}
\centering
\includegraphics[width=1.0\linewidth]{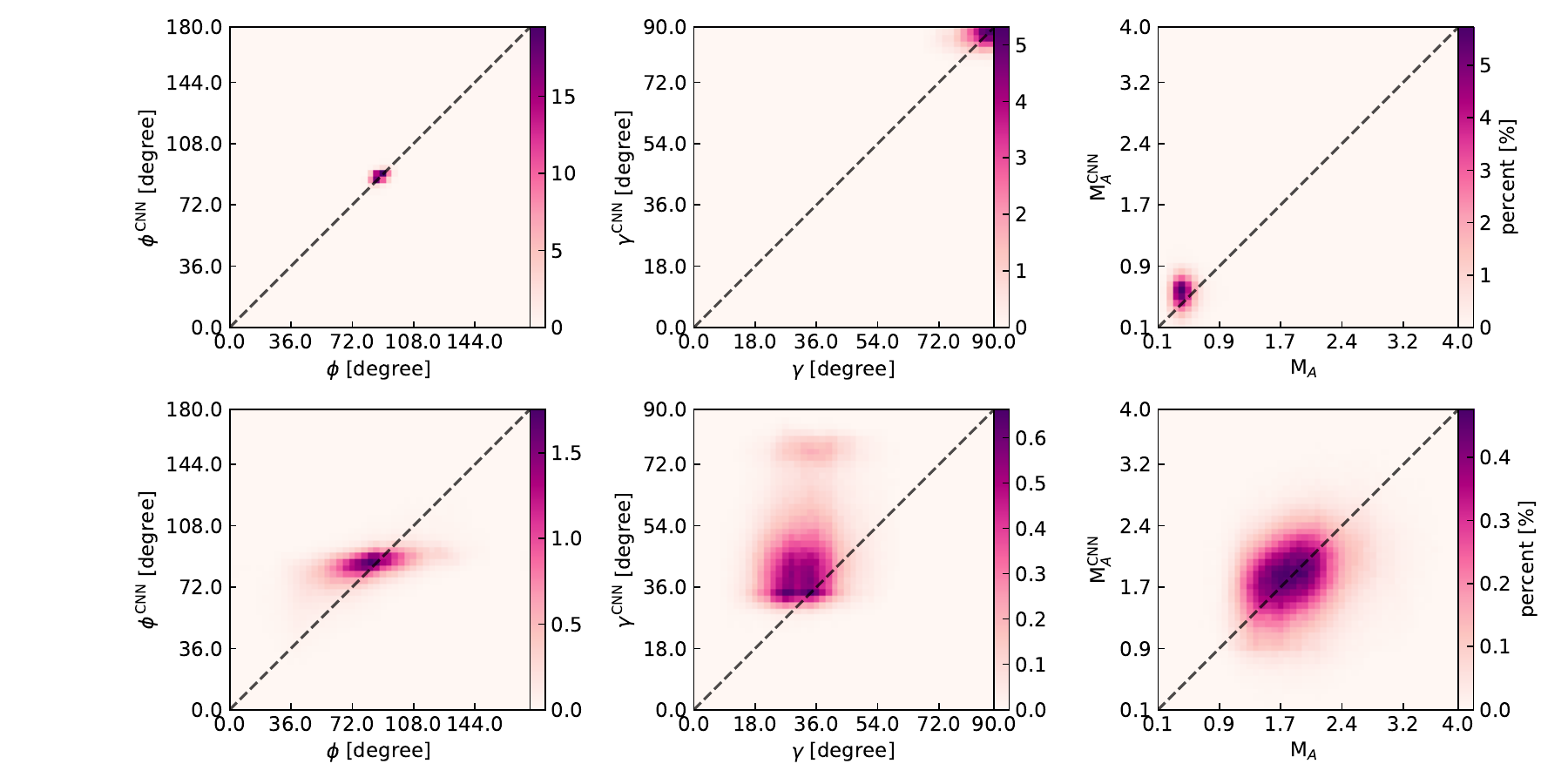}
        \caption{Same as Fig.~\ref{fig:2Dhist_13}, but for C$^{18}$O.}
    \label{fig:2Dhist_18}
\end{figure*}

Figs.~\ref{fig:2Dhist_13} and \ref{fig:2Dhist_18} present 2D histograms illustrating the correspondence between CNN predictions—$\phi^{\rm CNN}$, $\gamma^{\rm CNN}$, and M$_A^{\rm CNN}$—and actual values obtained from two test simulations, A2 and A6. In sub-Alfv\'enic cases for both $^{13}$CO and C$^{18}$O molecules, we observe a close alignment between the CNN predictions and the real values. The scatter of the predictions, which includes $\phi^{\rm CNN}$, $\gamma^{\rm CNN}$, and M$_A^{\rm CNN}$, demonstrates a small deviation from the actual values, tightly congregating near the one-to-one reference line. This minimal deviation suggests that the CNN model offers a high degree of accuracy and reliability when operating under sub-Alfvé\'enic conditions.

However, the scenario is a bit different in super-Alfv\'enic cases. Here, the scatter is noticeably more widespread, indicating that deviations from the real values increase in these conditions. The $\phi^{\rm CNN}$ predictions, in particular, show a tendency for both overestimation and underestimation. In contrast, the $\gamma^{\rm CNN}$ predictions are primarily characterized by overestimations, a trend that is especially prominent in cases involving C$^{18}$O molecules. Meanwhile, the scatter related to the M$_A^{\rm CNN}$ predictions is distributed more uniformly around the reference line.

This suggests predicting the 3D magnetic field under super-Alfv\'enic conditions is more challenging with higher uncertainty. In these environments, the magnetic field exerts a weaker influence, and the turbulence observed more closely resembles that of hydrodynamic turbulence, thereby complicating the prediction process. Enhancing prediction accuracy is feasible through two strategies. First, 
it is possible to further refine and optimize the CNN model to improve its adaptability and responsiveness to the unique features of super-Alfv\'enic MHD turbulence. For instance, \cite{2019ApJ...882L..12P} put forth a CNN model designed specifically to differentiate between sub-Alfv\'enic and super-Alfv\'enic turbulence. This model, with its specialized focus, offers a promising avenue for enhancing the accuracy of predictions in super-Alfv\'enic environments. Second, enrich the data set to train the CNN model. By incorporating a broader and more diverse range of images, the model can be exposed to a wider array of scenarios and conditions, thereby reducing uncertainty and improving its ability to make accurate predictions across different environments and conditions.

\begin{figure*}
\centering
\includegraphics[width=1.0\linewidth]{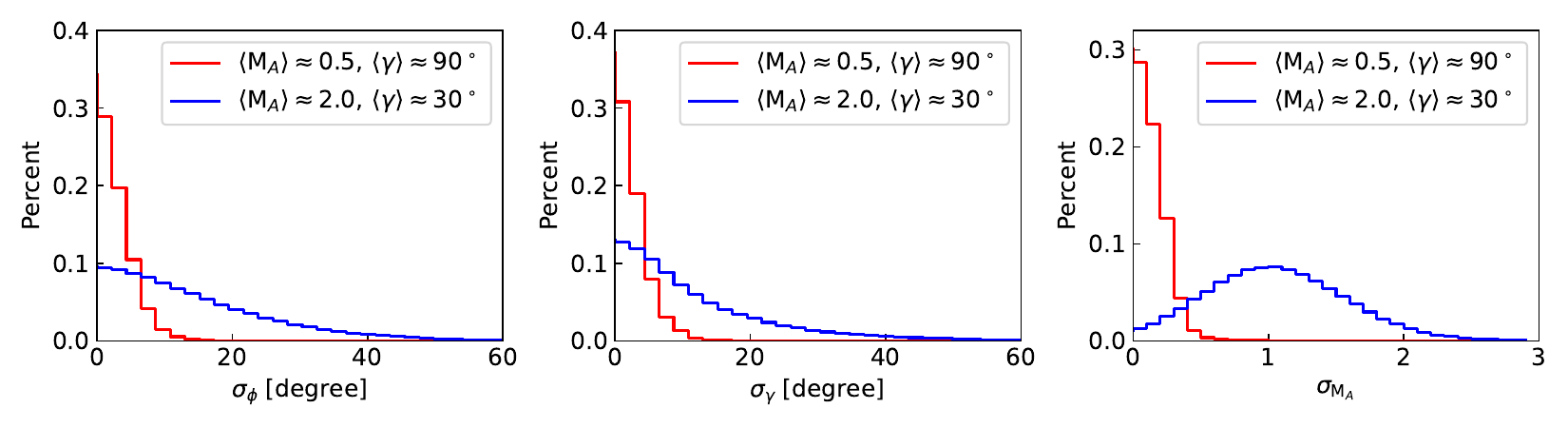}
        \caption{Histograms of difference in CNN-predicted $\phi^{\rm CNN}$ (left), $\gamma^{\rm CNN}$ (middle), and M$_A^{\rm CNN}$ (right) and the actual values in simulations using $^{13}$CO. }
    \label{fig:13sigma}
\end{figure*}

\begin{figure*}
\centering
\includegraphics[width=1.0\linewidth]{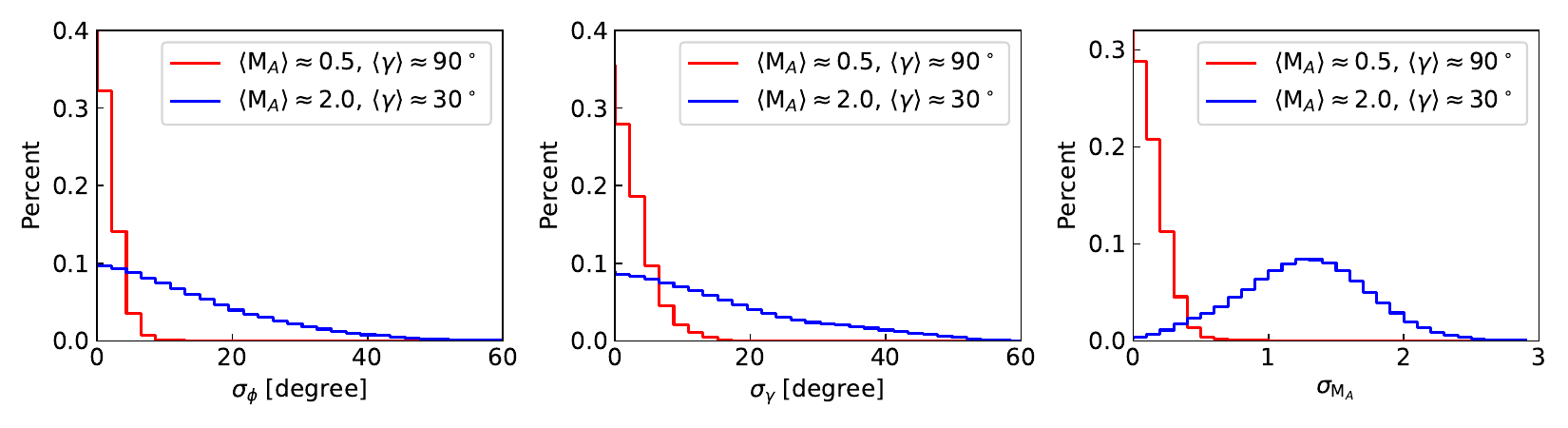}
        \caption{Same as Fig.~\ref{fig:13sigma}, but for C$^{18}$O.}
    \label{fig:18sigma}
\end{figure*}

\begin{figure*}
\centering
\includegraphics[width=0.99\linewidth]{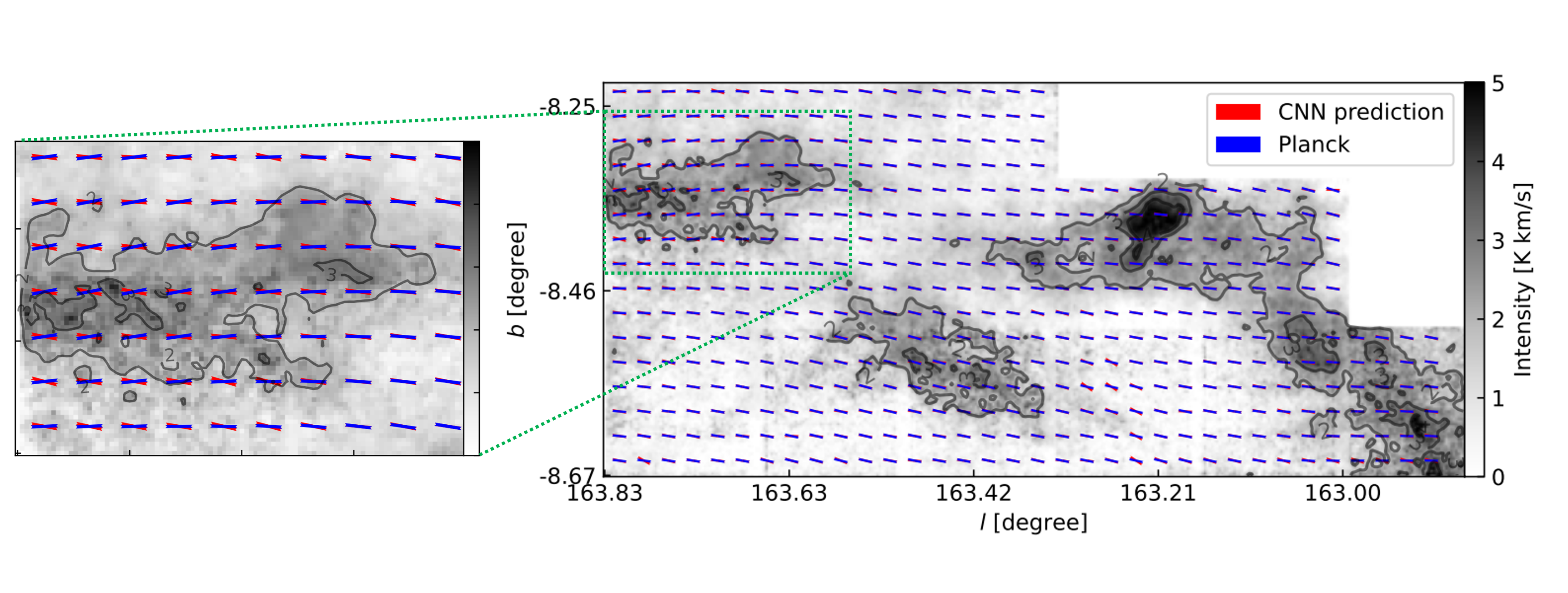}
        \caption{Comparison of the POS magnetic fields predicted by CNN-$^{13}$CO (red segment) for the L1478 cloud and inferred from Planck polarization (blue segment). The background image is the integrated $^{13}$CO intensity map. 
        %{\bf AL put zoom in sub-panel to show that there are two types of vectors that nearly coiside, but not identical.}
        }
    \label{fig:L1478_phi}
\end{figure*}

\begin{figure*}
	\centering
	\includegraphics[width=0.99\linewidth]{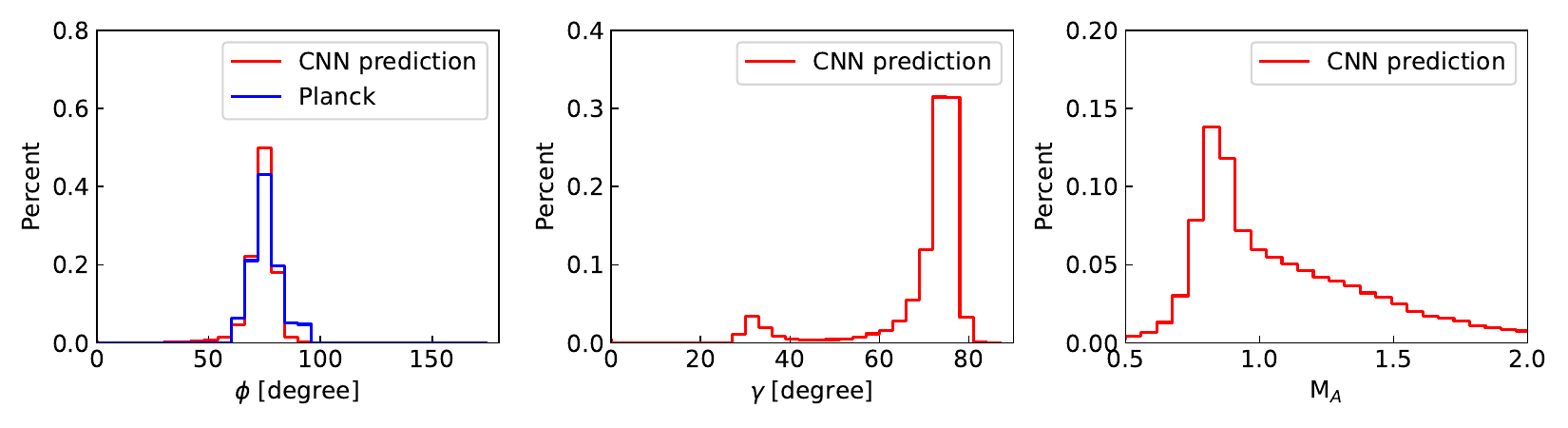}
	\caption{Histograms of CNN-predicted (as well as Planck measured) $\phi^{\rm CNN}$ (left), defined east from the north, $\gamma^{\rm CNN}$ (middle), and M$_A^{\rm CNN}$ (right).}
	\label{fig:L1478_hist}
\end{figure*}

\begin{figure*}
\centering
\includegraphics[width=0.5\linewidth]{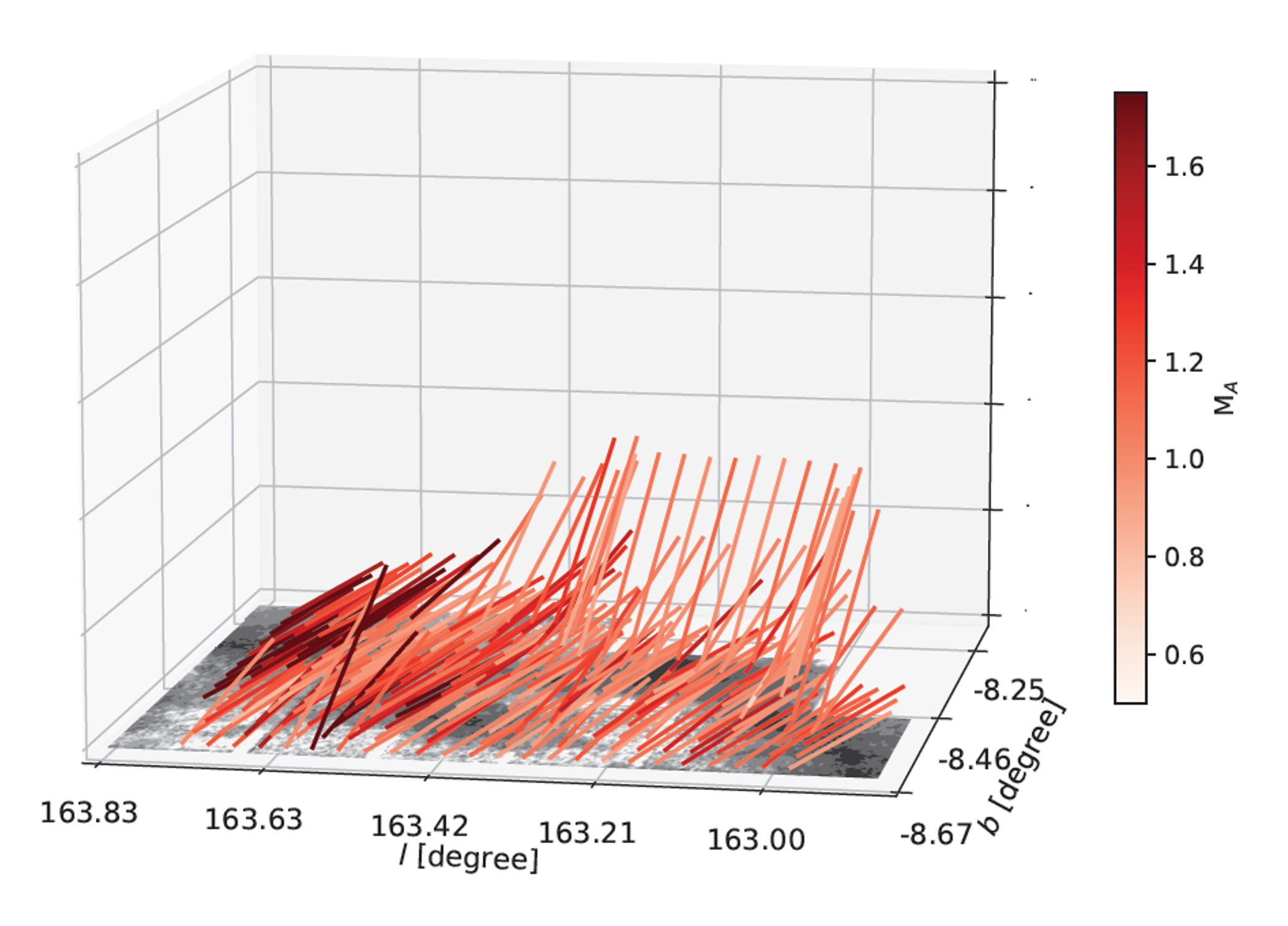}
        \caption{An visulization of the CNN-predicted 3D magnetic fields using $^{13}$CO for the L1478 cloud. Each magnetic field segment is constructed by the position angle of the POS magnetic field (i.e., $\phi$) and the inclination angle $\gamma$. Note that the magnetic field obtained is the projection along the LOS and averaged over 12$\times$12 pixels for visualization purposes. The third axis of the LOS is for 3D visualization purposes and does not provide distance information here. The total intensity map $I$ is placed on the POS, i.e., the $l-b$ plane.}
    \label{fig:L1478_3D}
\end{figure*}
Figs.~\ref{fig:13sigma} and \ref{fig:18sigma} plot the histograms of the deviation between the CNN-predicted and the actual 3D magnetic field. We calculate the absolute difference between $\phi^{\rm CNN}$ and $\phi$, between $\gamma^{\rm CNN}$ and $\gamma$, and between M$_A^{\rm CNN}$ and M$_A$, respectively. These differences are denoted as $\sigma_\phi$, $\sigma_\gamma$, and $\sigma_{{\rm M}_A}$. In the sub-Alfvénic scenarios, we observed that the distributions of $\sigma_\phi$ and $\sigma_\gamma$ are relatively condensed, primarily falling within the 0 to 20$^\circ$ range. This concentration indicates a close alignment between the CNN predictions and the actual values in sub-Alfv\'enic environments, suggesting that the CNN model performs with high precision in these conditions. However, as $\langle {\rm M}_A\rangle$ increases, the distributions of $\sigma_\phi$ and $\sigma_\gamma$ broaden, spanning a more extensive range from 0 to 60$^\circ$. This dispersion is indicative of larger deviations between predicted and actual values under these conditions, implying that the CNN model may face challenges in accurately capturing the magnetic field dynamics when $\langle {\rm M}_A\rangle$ increases.

Examining specific molecules, for $^{13}$CO under sub-Alfv\'enic conditions, the median deviation values are relatively low: 
$\sigma_\phi=3.26^\circ$, $\sigma_\gamma=2.98^\circ$, and $\sigma_{{\rm M}_A}=0.16$. In contrast, under super-Alfv\'enic conditions, these values increase to 12.32$^\circ$, 9.08$^\circ$, and 1.1, respectively, highlighting an increase in prediction deviation as the environment transitions from sub- to super-Alfv\'enic. Similarly, for C$^{18}$O, the median deviation values are 2.22$^\circ$, 3.20$^\circ$, and 0.16 under sub-Alfv\'enic conditions and 12.08$^\circ$, 13.60$^\circ$, and 1.36 under super-Alfv\'enic scenarios, underlining a consistent trend of increased deviation in super-Alfv\'enic environments across different molecules.

\subsection{Observational prediction}
For the observational tests, our target is the nearby L1478 cloud. We utilized $^{13}$CO spectral line from a previous study \cite{2021ApJ...908...76L}. The data has a beam resolution of 38" and was regrid to a pixel resolution of 10", while achieving a velocity resolution of 0.3 km s$^{-1}$. The 1D velocity dispersion $\sigma_v$ of the $^{13}$CO line was reported within the range of $0.40-0.70~{\rm km~s^{-1}}$ \citep{2021ApJ...908...76L}. Assuming an isotropic velocity dispersion in 3D and uniform temperature of 10 K (corresponding to an isothermal sound speed of $c_s\sim0.187~{\rm km ~s^{-1}}$, see \citealt{HLS21}), we find the sonic Mach number M$_s=\sqrt{3}\sigma_v/c_s$ ranges from 3.69 to 6.45, falling into the parameter regimes in our numerical simulations. With these refined data, we applied our adeptly trained CNN model to the $^{13}$CO channel map, aiming to predict the key 3D magnetic field parameters, denoted as $\phi^{\rm CNN},~\gamma^{\rm CNN},~{\rm M}_A^{\rm CNN}$.

For the purpose of validating the results yielded through our CNN application, we engaged in a comparative analysis with POS magnetic field orientations as predicted through Planck 353 GHz polarization data. The data harnessed for this comparative process was drawn from the third Public Data Release (DR3), provided by Planck's High-Frequency Instrument \citep{2020A&A...641A...3P}. The POS magnetic field orientation was inferred from Stokes parameters $Q$ and $U$ converted to IAU convention from HEALPix using the equation: $\phi^{\rm Planck} = \frac{1}{2}\tan^{-1}(-U,Q) + \pi/2$. To enhance the signal-to-noise ratio, we smoothed the Stokes parameter maps from an angular resolution of 5$'$ to 10$'$ using a Gaussian kernel.

As presented in Fig.~\ref{fig:L1478_phi}, a remarkable alignment between the magnetic field orientations as predicted by both the CNN model and the Planck polarization data is observed, while we notice the difference is apparent in the northeast clump (see the zoom-in plot in Fig.~\ref{fig:L1478_phi}). To quantify the agreement between CNN-prediction and polarization, we utilize the Alignment Measure (AM; \citealt{GL17}), expressed as:
\begin{equation}
\begin{aligned}
{\rm AM} = \langle\cos(2\theta_{\rm r})\rangle,
\end{aligned}
\end{equation}
here $\theta_{\rm r}$ is the relative angle between the two measurements. 
An AM  value of $\approx0.94$ confirms the CNN-prediction has an excellent agreement with Planck polarization \footnote{AM = 1 implies a perfect parallel alignment, while -1 indicates perpendicularity.}, corresponding to an overall deviation of $\approx10^\circ$. 

%{\bf AL This alignment measure corresponds to the dispersion of angles $\approx ...$. The accuracy of the Plank polarization mapping of magnetic field directions is $\approx ...$}

A noteworthy advantage of our CNN model over traditional polarization methodologies is its ability to trace the 3D magnetic fields. This is achieved through the model’s predictions regarding $\gamma$ and M$_A$. These predictions are summarised in histograms within Fig.\ref{fig:L1478_hist}. According to the histograms, the median $\gamma$ and M$_A$ of the L1478 cloud are estimated at $\approx76^\circ$ and $\approx1.07$, respectively. These measurements suggest that the L1478 is a trans-Alfv\'enic could. In this state, there is an equilibrium between magnetic and turbulent kinetic energies within the cloud. The parameters derived from the CNN application have been instrumental in creating the first-ever 3D magnetic field map for L1478, which can be viewed in Fig.~\ref{fig:L1478_3D}.

\section{Discussion}
\label{sec:dis}
\subsection{Comparison with earlier studies}

The realm of exploring magnetic fields within the ISM through CNN is experiencing swift advancements. As a pilot study presented by \cite{2023ApJ...942...95X}, the Convolutional Approach to Structure Identification-3D (CASI-3D) model was employed to map the 2D POS magnetic field orientation. This is achieved similarly by using the velocity channel maps obtained from spectroscopic observations. The underlying physics principle is still founded on the anisotropic MHD turbulence. The training process underpinning this approach uses the emission lines of $^{12}$CO and $^{13}$CO (J = 1 - 0), generated through the RADMC-3D code \citep{2012ascl.soft02015D}.

In this study, we introduce a new CNN model. This advanced model is designed with the aim of predicting not merely the orientation $\phi$ of the POS magnetic field but extends to encompass the angle of field inclination, $\gamma$, as well as the total Alfv\'en Mach number M$_A$. This approach allows the construction of 3D magnetic field vectors. For training the CNN model, we have utilized emission lines from $^{13}$CO and C$^{18}$O (J = 1 - 0), with data generated from the SPARX code \citep{2019ApJ...873...16H}.

We quantify the uncertainty of our CNN-predicted $\phi$ and $\gamma$. We found that the median value and the dispersion of uncertainty for C$^{18}$O are approximately $\sim2.22^\circ$ and $\sim3.20^\circ$ under sub-Alfv'enic conditions ($\langle {\rm M}_A\rangle\approx0.5$). These values shift to $\sim12.08^\circ$ and $\sim13.60^\circ$ under super-Alfv\'enic conditions ($\langle {\rm M}_A\rangle\approx2.0$). When compared to the CASI-3D model, our CNN model demonstrates higher accuracy, as CASI-3D exhibits a median uncertainty of $\sim6.2^\circ$ and $\sim18.4^\circ$ under comparable sub-Alfv\'enic and super-Alfv\'enic conditions, respectively. Through the application of our CNN model to the L1478 molecular cloud, we successfully constructed the first 3D magnetic field map. The corresponding CNN-predicted POS magnetic field orientation shows remarkable alignment with that inferred from Planck 353 GHz polarization data.

It is crucial to acknowledge that despite the differences inherent between the CNN models used by \cite{2023ApJ...942...95X} and this study, the fundamental concept of utilizing spectroscopic channel maps for magnetic field investigation remains the same: (1) the intensity distribution observable in thin channel maps is predominantly influenced by turbulent velocity statistics, as outlined in \citep{LP00,2016MNRAS.461.1227K,2023MNRAS.tmp.1894H}; and (2) these channel maps capture the anisotropy intrinsic to MHD turbulence, thereby revealing the orientation of the POS magnetic field \citep{LY18a,2023MNRAS.tmp.1894H}. A crucial insight was provided by \cite{HLX21a}, highlighting that the degree of anisotropy in channel maps, as well as the magnetic field topology, is regulated by both the $\gamma$ and the M$_A$. These are parameters that can be extracted efficiently using the CNN approach \footnote{Note that while the anisotropy and magnetic field topology, that are sensitive to $\gamma$ and the M$_A$, are the most apparent features in channel maps, it is also possible the CNN extracts additional features to facilitate the prediction.}. Thus, drawing upon these foundational theoretical studies, we propose the use of the CNN model as an efficient tool for tracing 3D magnetic fields, providing convincing physical reasons for interpreting its feasibility.

\subsection{Synergy with other methods}
Our newly proposed CNN model stands as a powerful complement to existing methodologies in the field. One notable technique, which involves utilizing polarized dust emission, has proven effective in tracing the 3D magnetic field orientation within diffuse clouds, where dust grains are perfectly aligned with magnetic fields  \citep{2019MNRAS.485.3499C,2023MNRAS.519.3736H,2023MNRAS.524.4431H}. However, this technique may encounter limitations within dense cloud environments – for example, those observed through tracing by C$^{18}$O – where dust grains might not maintain perfect alignment \citep{Lazarian07,BG15}. This loss of alignment, resulting in a phenomenon known as the polarization hole \citep{2019MNRAS.482.2697S,2019ApJ...880...27P,2021ApJ...908..218H}, introduces uncertainties when tracing 3D magnetic fields through polarized dust emission techniques.

Unlike these traditional approaches, the CNN approach remains immune to the effects of the polarization hole. When the CNN model is supplied with emission lines from dense tracers like C$^{18}$O, HNC, and NH$_3$, it proves highly adept at probing the 3D magnetic fields present within dense clouds effectively. Nonetheless, it's important to consider that within these dense cloud environments, the forces of self-gravity can become a significant factor. This gravitational influence might induce alterations in the anisotropy observed within channel maps \citep{HLY20}. Therefore, it becomes imperative to input the CNN model with carefully selected numerical simulations before applying it to observational data to ensure accurate and reliable results.

Furthermore, it should be noted that the inclination angle predicted by the CNN model is inherently limited to the range of [0, 90$^\circ$]. This limitation arises because the anisotropy within channel maps alone cannot definitively discern whether the magnetic field is oriented towards or away from the observer. However, recent advancements in the field, particularly in Faraday rotation measurements within molecular clouds \citep{2019A&A...632A..68T,2022A&A...660A..97T}, offer promising avenues to resolve this degeneracy.

Another relevant method worth discussing is the Velocity Gradient Technique (VGT; \citealt{GL17,LY18a,HYL18}). Like our proposed CNN approach, the VGT is a technique that traces magnetic fields using spectroscopic observations. Importantly, both the CNN approach and VGT share a foundational physical principle: they rely on the anisotropy of MHD turbulence observed within thin channel maps. With VGT having undergone extensive and rigorous testing \citep{Hu19a,2020MNRAS.496.2868L,HLS21,2022MNRAS.510.4952L,2022A&A...658A..90A,2023MNRAS.519.1068L,2023MNRAS.524.2379H,2023ApJ...946....8T,2023MNRAS.523.1853S}, it is established as an excellent benchmark for evaluating the accuracy of CNN models, especially in situations where polarization measurements are not readily available. This benchmarking is crucial when CNNs are deployed for tracing 3D Galactic Magnetic Fields, highlighting the important comparative and complementary roles these techniques play in advancing our understanding of magnetic fields in various astrophysical contexts.

\subsection{Prospects of the CNN method}
In the present study, we introduced a CNN model adept at predicting 3D magnetic fields within molecular clouds, utilizing spectroscopic observations of molecular gas. However, the potential applications of this CNN method extend far beyond, encompassing various astrophysical environments and contexts, including neutral hydrogen (HI) regions, ionized gas, the Central Molecular Zone (CMZ), external galaxies, and supernova remnants. In the following sections, we outline several promising applications of this methodology.

\subsubsection{3D Galactic Magnetic Fields}
A deep and comprehensive understanding of the 3D Galactic Magnetic Field (GMF; \citealt{2012ApJ...761L..11J}) is paramount for addressing a host of astrophysical inquiries. These include identifying the origins of ultra-high energy cosmic rays \citep{2014CRPhy..15..339F,2019JCAP...05..004F} and refining models of Galactic foreground polarization \citep{2015PhRvD..91h1303K,2016A&A...594A..25P}.

Recent research indicates that thin channel maps of HI successfully capture the anisotropy inherent in MHD turbulence \citep{LP00,LY18a,2020MNRAS.496.2868L,2023MNRAS.tmp.1894H}. Consequently, applying the CNN to HI channel maps constitutes a viable strategy for mapping 3D GMFs. Past efforts aimed at modeling the foreground polarization with HI primarily focused on mapping the POS magnetic field orientation \citep{2019ApJ...887..136C,2020MNRAS.496.2868L,2020ApJ...888...96H}. These endeavors largely neglected the crucial depolarization factor, the inclination angle. However, the advent of sophisticated multi-phase HI simulations \citep{2021arXiv211106845H} has made it possible to train the CNN model for accurate predictions of 3D GMFs, yielding more realistic models of the foreground polarization.

Our primary goal in this paper is to explore the magnetic fields of molecular clouds, for which the isothermal approximation is applicable. 
Multi-phase HI requires separate training of the neural network. For multi-phase HI where cooling and heating play a significant role, our general approach remains valid: intensity features/striations within channel maps continue to elongate along the POS magnetic field orientation. This is supported by several studies \cite{LY18a,2019ApJ...887..136C,2020MNRAS.496.2868L,2020ApJ...888...96H,2023MNRAS.tmp.1894H}. These intensity features/striations are also regulated by the Alfv\'en Mach number (M$_A$) and the projection effect associated with the inclination angle. However, additional physics, such as thermal instability, could modify the observed anisotropy, for instance, potentially leading to a smaller aspect ratio \citep{2023MNRAS.521..230H}. %While these factors do not compromise the probe of the POS magnetic field, multi-phase simulations could offer more reliable estimations of M$_A$ and the inclination angle, reducing uncertainty. 
The corresponding study employing our approach for multiphase HI will be provided elsewhere.

\subsubsection{3D Magnetic Fields in CMZ and external galaxies}
Understanding the magnetic fields within cold molecular gas is essential for deciphering the processes of formation and fueling of Seyfert nuclei. Recent measurements of magnetic fields within the CMZ and in other Seyfert galaxies have been conducted using various techniques. These include far-infrared polarization observations from instruments like SOFIA/HAWC+ \citep{2021ApJ...923..150L}, JCMT \citep{2021MNRAS.505..684P}, and ALMA \citep{2020ApJ...893...33L}, as well as employing the VGT \citep{2022MNRAS.511..829H,2022ApJ...941...92H,2023MNRAS.519.1068L}. However, these approaches primarily yield the POS magnetic field orientation, falling short of providing a comprehensive 3D perspective. Nevertheless, the successful application of VGT confirms the viability of using anisotropy in molecular emission channel maps as a tracer for magnetic fields in these environments. For instance,  \cite{2022MNRAS.511..829H} derived a POS magnetic field map surrounding Sgr A* using the [Ne II] emission line and Paschen-$\alpha$ image observed with the Hubble Space Telescope (HST). Given these advances, extending the CNN methodology to incorporate optical/near-infrared observations from instruments like the HST and the James Webb Space Telescope is a feasible and promising approach for predicting 3D magnetic fields in both the Galactic center and external galaxies.

\subsection{Obtaining the full 3D magnetic field vector}
3D magnetic fields, encompassing both orientation and strength, play a pivotal role in comprehending key astrophysical phenomena. These include processes such as star formation \citep{1965QJRAS...6..265M,MK04,MO07,2012ApJ...757..154L,2012ApJ...761..156F,HLS21}, the effects of stellar feedback \citep{2022MNRAS.515.1026P,2023arXiv230904173L}, as well as the acceleration and propagation of cosmic rays \citep{1949PhRv...75.1169F,1966ApJ...146..480J,2002PhRvL..89B1102Y,2013ApJ...779..140X,2020ApJ...894...63X,2021arXiv211115066H,2022FrASS...9.0900B,2023ApJ...956...63L}. Traditionally, to obtain the strength of these fields, the Davis-Chandrasekhar-Fermi (DCF) method is employed, which typically combines dust polarimetry with spectroscopic observations (see \citealt{1951PhRv...81..890D,1953ApJ...118..113C}). However, this often proves insufficient. the DCF method gives only the POS magnetic field strength, while the component along the LOS is missing. Other limitations of the DCF method have also been thoroughly dissected in the literature \citep{2021A&A...656A.118S,2022ApJ...935...77L,2022MNRAS.514.1575C,2022ApJ...925...30L}.

In light of this, an alternative approach has been proposed: the use of the Alfv\'en Mach number M$_A$ with the sonic Mach number M$_s$ to derive the magnetic field's strength \citep{2020arXiv200207996L}. This method, aptly termed MM2, 
can be used to obtain the total strength, particularly since the vital term M$_A$ is readily available with the CNN approach proposed in this study. The sonic Mach number 
M$_s$ can be procured either directly via spectroscopic line broadening or by leveraging a CNN approach similar to our current study. Coupled with the 3D magnetic field orientation, this equips us with the necessary tools to construct a 3D magnetic field vector.

\section{Summary}
\label{sec:con}
In this study, a CNN model was designed for the intricate task of probing 3D magnetic fields within molecular clouds. This model is not confined to determining the POS magnetic field orientation but extends its capabilities to accurately ascertain the field's inclination angle and the total Alfv\'en Mach number, offering a more comprehensive understanding of the magnetic field in the observed regions. We summarize our major results below:
\begin{enumerate}
    \item We developed a CNN model for probing the 3D magnetic fields, including the POS magnetic field orientation, inclination angle, and total Alfv\'en Mach number.
    \item The CNN model was trained using synthetic $^{13}$CO and C$^{18}$O (J = 1 - 0) emission lines, encompassing a range of conditions from sub-Alfv\'enic to super-Alfv\'enic. We quantified the uncertainties associated with the trained CNN model's predictions. Our findings revealed that the uncertainties are less than $5^\circ$ for both $\phi$ and $\gamma$, and are smaller than $0.2$ for M$_A$ under sub-Alfv\'enic conditions (with M$_A\approx0.5$). Under super-Alfv\'enic conditions (with M$_A\approx2.0$), the uncertainties increased slightly but remained below $15^\circ$ for $\phi$ and $\gamma$, and were around $1.5$ for M$_A$.
    \item We implemented our trained CNN model to analyze the molecular cloud L1478. The CNN-predicted POS magnetic field orientation exhibited remarkable agreement with orientations inferred from Planck 353 GHz polarization data, with a marginal global difference of approximately $10^\circ$.
    \item This study facilitated the construction of the first 3D magnetic field map for the L1478 cloud. Through our analysis, we found that the cloud's global inclination angle is approximately $76^\circ$, while the global total Alfv\'en Mach number is close to $1.07$. 
    \item  We discussed the potential applications and future prospects of the CNN approach. Particularly, we discussed the feasibility and potential of utilizing the CNN model for predicting 3D Galactic Magnetic Fields. We also considered its application for understanding 3D magnetic fields in the CMZ and external galaxies.
\end{enumerate}

\section*{Acknowledgements}
Y.H. and A.L. acknowledge the support of NASA ATP AAH7546, NSF grants AST 2307840, and ALMA SOSPADA-016. Financial support for this work was provided by NASA through award 09\_0231 issued by the Universities Space Research Association, Inc. (USRA). This work used SDSC Expanse CPU at SDSC through allocations PHY230032, PHY230033, PHY230091, and PHY230105 from the Advanced Cyberinfrastructure Coordination Ecosystem: Services \& Support (ACCESS) program, which is supported by National Science Foundation grants \#2138259, \#2138286, \#2138307, \#2137603, and \#2138296. Y.H. acknowledges the very kind computational and technical support of Bowen Cao.

%%%%%%%%%%%%%%%%%%%%%%%%%%%%%%%%%%%%%%%%%%%%%%%%%%
\section*{Data Availability}
The data underlying this article will be shared on reasonable request to the corresponding author. 


\bibliographystyle{mnras}
\bibliography{example} % if your bibtex file is called example.bib

%\begin{thebibliography}{99}
%\end{thebibliography}

%%%%%%%%%%%%%%%%%%%%%%%%%%%%%%%%%%%%%%%%%%%%%%%%%%

%%%%%%%%%%%%%%%%% APPENDICES %%%%%%%%%%%%%%%%%%%%%
%\newpage
\appendix
%%%%%%%%%%%%%%%%%%%%%%%%%%%%%%%%%%%%%%%%%%%%%%%%%%
\section{Anisotropy's degeneracy on $\gamma$ and M$_A$}
\label{appendix.A}
\begin{figure*}
\includegraphics[width=0.99\linewidth]{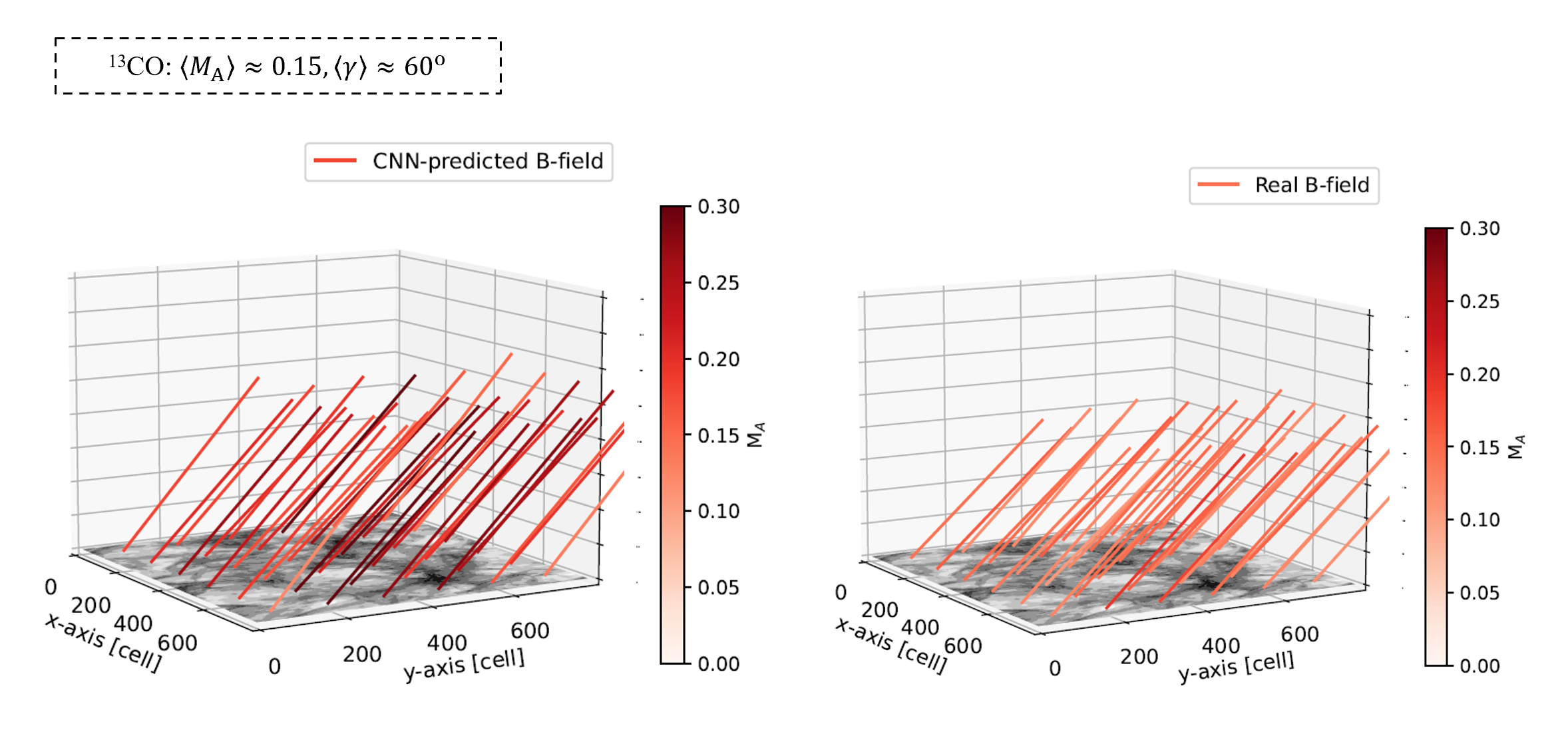}
        \caption{Same as Fig.~\ref{fig:3D_13co}, but for a different simulation with $\langle {\rm M}_A\rangle\approx0.15$ and $\langle \gamma\rangle\approx60^\circ$.}
    \label{fig:3D_13cob}
\end{figure*}
As we discussed in \S~\ref{sec:theory},  the degree of observed anisotropy is intricately modulated by both $\gamma$ and M$_A$
, resulting in a degeneracy. However, it is expected that the CNN model has the capacity to extract additional information, such as changes in the magnetic field topology, to simultaneously determine both $\gamma$ and M$_A$. Figs.~\ref{fig:3D_13co} and ~\ref{fig:3D_13cob} present a test for this.

The simulation deployed in Fig.~\ref{fig:3D_13co} maintains average conditions of $\langle {\rm M}_A\rangle\approx0.5$ and $\langle \gamma\rangle\approx90^\circ$, while Fig.~\ref{fig:3D_13cob} operates under $\langle {\rm M}_A\rangle\approx0.15$ and $\langle \gamma\rangle\approx60^\circ$. Both simulations are expected to exhibit comparable degrees of anisotropy on the POS. We can see in both cases, the CNN model adeptly reconstructs the 3D magnetic fields. This reconstruction is achieved through accurate predictions of both the POS magnetic field orientation and the value of $\gamma$. However, the CNN-predicted M$_A$ is approximately 0.1 to 0.2 larger than the actual M$_A$, indicating an overestimation. This testifies that the CNN model is capable of solving the anisotropy degree's degeneracy. The overestimation in M$_A$ might be addressed through possible solutions, as discussed in \S~\ref{sec:result}.

\section{$^{13}$CO emission line}

%\begin{figure}
%\includegraphics[width=0.99\linewidth]%{figure/spectrum.pdf}
%        \caption{The spectrum of $^{13}$CO averaged over the entire L1478 cloud.}
%    \label{fig:spectrum}
%\end{figure}

%Fig.~\ref{fig:spectrum} presents the spectrum of $^{13}$CO averaged over the entire L1478 cloud. We fitted the spectrum with a Gaussian distribution and obtained the velocity dispersion $\sigma_v\sim0.94$ km/s.
% Don't change these lines
\bsp	% typesetting comment
\label{lastpage}
\end{document}